\documentclass[twocolumn,amsmath,amssymb,longbibliography,aps, prb,floatfix]{revtex4-2}

\usepackage{graphicx}
\usepackage{dcolumn}
\usepackage{bm}
\usepackage[dvipsnames]{xcolor}
\usepackage{hyperref}

\newcommand{\vs}{\textbf{s}}
\usepackage[compat=1.1.0]{tikz-feynman} 

\usepackage{amsmath}
\allowdisplaybreaks 

\newcommand{\vk}{\mathbf{k}}
\newcommand{\vq}{\mathbf{q}}
\newcommand{\vp}{\mathbf{p}}
\newcommand{\vd}{\mathbf{d}}
\newcommand{\vB}{\mathbf{B}}
\newcommand{\nb}{NbSe$_2$}

\begin{document}


\title{Unconventional pairing in Ising superconductors: Application to monolayer NbSe$_2$}


\author{Subhojit Roy}
\affiliation{Department of Physics, Indian Institute of Technology Madras, Chennai, 600036, India}
\affiliation{Center for Atomistic Modelling and Materials Design, IIT Madras, Chennai 600036, India}
\affiliation{Quantum Centers in Diamond and Emergent Materials (QCenDiem)-Group, IIT Madras, Chennai, 600036 India}
\author{Andreas Kreisel}
\author{Brian M. Andersen}
\affiliation{Niels Bohr Institute, University of Copenhagen, DK-2100 Copenhagen, Denmark}
\author{Shantanu Mukherjee}
\email[Corresponding Author: ]{shantanu@iitm.ac.in}
\affiliation{Department of Physics, Indian Institute of Technology Madras, Chennai, 600036, India}
\affiliation{Center for Atomistic Modelling and Materials Design, IIT Madras, Chennai 600036, India}
\affiliation{Quantum Centers in Diamond and Emergent Materials (QCenDiem)-Group, IIT Madras, Chennai, 600036 India}

\begin{abstract}
The presence of a non-centrosymmetric crystal structure and in-plane mirror symmetry allows an Ising spin-orbit coupling to form in some two-dimensional materials. Examples include transition metal dichalcogenide superconductors like monolayer NbSe$_2$, MoS$_2$, TaS$_2$, and PbTe$_2$, where a nontrivial nature of the superconducting state is currently being explored. In this study, we develop a microscopic formalism for Ising superconductors that captures the superconducting instability arising from a momentum-dependent spin- and charge-fluctuation-mediated pairing interaction. We apply our pairing model to the electronic structure of monolayer \nb{}, where first-principles calculations reveal the presence of strong paramagnetic fluctuations. Our calculations provide a quantitative measure of the mixing between the even- and odd-parity superconducting states and its variation with Coulomb interaction. Further, numerical analysis in the presence of an external Zeeman field reveals the role of Ising spin-orbit coupling and mixing of odd-parity superconducting state in influencing the low-temperature enhancement of the critical magnetic field.
\end{abstract}

\maketitle

\section{\label{sec:level1}Introduction}

Monolayer transition metal dichalcogenides (TMDs) have drawn a lot of attention owing to their ability to host emergent phenomena such as superconductivity~\cite{frindt1972superconductivity}, charge density wave (CDW) order~\cite{leroux2015strong,guster2019coexistence,xi2015strongly,ugeda2016characterization,bianco2020weak}, magnetism~\cite{bonilla2018strong} and topological properties~\cite{he2018magnetic}. They hold potential for applications across a spectrum of electronic, optical, spintronics, and valleytronics devices~\cite{splendiani2010emerging,radisavljevic2011single,sundaram2013electroluminescence,lopez2013ultrasensitive,rycerz2007valley,cao2011mos_2,mak2012control,zeng2012valley,husain2018spin,husain2020emergence}. 

\begin{figure*}
    \includegraphics[width=\linewidth]{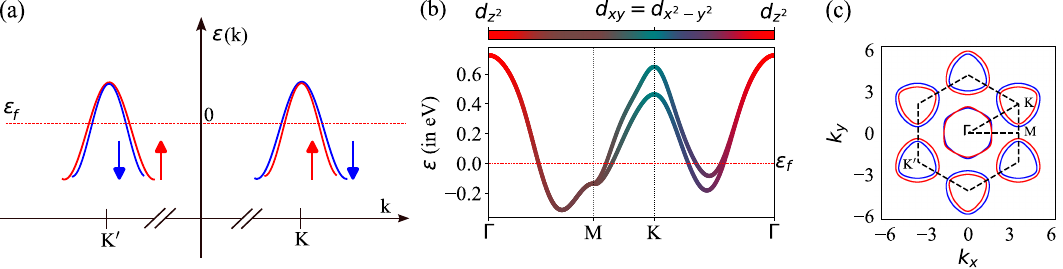} 
   \caption{(a) Schematic representation of spin splitting of bands in TMD monolayers at the $K$ and $K'$ points due to Ising SOC. (b) Low-energy orbital resolved electronic structure of monolayer \nb{}. (c) Spin-resolved Fermi surface in the presence of Ising SOC. The results shown in (b) and (c) are obtained form relativistic DFT calculations. 
   }
    \label{fig_ising_band_fs}
\end{figure*}

In monolayer TMDs (MX$_{2}$), the triangularly arranged transition metal (M) atoms are sandwiched between two layers of triangularly arranged chalcogenide atoms (X), giving rise to a 2D honeycomb lattice structure~\cite{bromley1972band,boker2001band}. One of the intriguing features of such monolayer TMDs is the lack of inversion symmetry, which leads to the formation of a large anti-symmetric spin-orbit coupling (SOC) and associated splitting of the energy bands. Generically, the fermiology consists of bands forming Fermi pockets close to the $K$ and $K'$ valleys, see Fig.~\ref{fig_ising_band_fs}(a). The presence of time-reversal symmetry further ensures that a spin-up state at the $K$ valley is degenerate with a spin-down state at the $K'$ valley. If the system additionally contains an in-plane mirror symmetry, the electronic spins are aligned out of the plane, and the SOC is termed Ising SOC. This kind of arrangement reduces the effects of in-plane Zeeman fields on the band structure~\cite{xi2016ising,saito2016superconductivity,dvir2018spectroscopy,liu2018interface,sohn2018unusual,de2018tuning}, and the corresponding superconducting system is dubbed an Ising superconductor~\cite{xing2017ising,wan2022observation,xi2016ising,dvir2018spectroscopy,mockli2018robust,fischer2023superconductivity,wickramaratne2021magnetism,de2018tuning,lu2015evidence,kim2017quasiparticle,saito2016superconductivity,liu2018interface}. Features such as the robustness of the superconducting state against in-plane magnetic fields owing to the Ising SOC, the mixing of odd- and even-parity superconducting states, and non-trivial topological properties make these Ising superconductors an exciting case study for understanding the role of SOC in 2D superconducting materials~\cite{Zhang_2021, LI2021100504, Wang2021, Wickramaratne_review}.

Monolayer \nb{} is a prime candidate where Ising superconductivity can be studied. Bulk \nb{} has been extensively investigated experimentally and the transition at $T_{\mathrm{c}}\sim 7$K into the superconducting state is believed to be driven by strong electron-phonon coupling \cite{Boaknin2003,valla2004quasiparticle, Noat2015} with a superconducting gap corresponding to an $s$-wave symmetry. On the other hand, as the number of Nb layers is reduced, the transition temperature drops, and in monolayer \nb{}, the superconducting state is observed only below $T_c\sim3$ K \cite{frindt1972superconductivity,staley2009electric,cao2015quality,ugeda2016characterization}.
This suppression of $T_{\mathrm{c}}$ from the bulk to monolayer can be due to the enhancement of thermally driven superconducting phase fluctuations and the weakening of the strength of Cooper pairing in 2D. An additional possible factor contributing to this weakening is the expected reduced screening of the Coulomb interactions.

The pairing mechanism in monolayer \nb{} is, however, not well understood. A number of proposals have considered electron-phonon mediated superconducting pairing~\cite{das2023electron,zheng2019electron,xi2015strongly,wang2023decisive,lian2018unveiling}, while other models consider pairing arising from purely repulsive  interactions~\cite{das2021quantitative,costa2022ising,hsu2017topological,shaffer2020crystalline,Ammar2023,Shaffer2023,Horhold_2023}. There are also proposals considering a combination of electron-phonon and spin-fluctuation mediated pairing~\cite{das2023electron}. First principal calculations have been utlized to study interplay between charge density wave (CDW) and superconducting phases in monolayer NbSe$_{2}$~\cite{zheng2019electron,wang2023decisive} where a conventional attractive electron-phonon superconducting pairing mechanism in the CDW state leads to an anisotropic s-wave gap over the Fermi surface. Other theoretical approaches have investigated the possibility of unconventional superconductivity from a repulsive pairing interaction that mixes even and odd parity superconducting states that correspond to sign changing gap structures that break the lattice point group symmetries. Within such repulsive interaction models there have been proposals for topological superconductivity in monolayer NbSe$_{2}$~\cite{shaffer2020crystalline,Ammar2023}, and models to demonstrate two-band nature of superconductivity in monolayer NbSe$_2$~\cite{Horhold_2023}. Other works have also explored the possibility of a combined role of electron phonon interaction and an unconventional repulsive pairing and proposed the possibility for a s-wave type superconducting ground state with a negligible gap on the $\Gamma$ centered pocket~\cite{das2023electron}.
A study of the electric field effect on superconductivity in monolayer \nb{} hints at a transition from a weak-coupling to a strong-coupling
superconductor as the single crystal is thinned to the atomic length scale~\cite{staley2009electric}. STM experiments indicate the presence of bosonic, undamped collective modes that could suggest the important role of spin fluctuations~\cite{wan2022observation}. The presence of strong spin fluctuations has also been supported by an enhanced paramagnetic susceptibility extracted from DFT calculations~\cite{das2023electron}. The possibility of an unconventional superconducting order has also been ascribed to studies of magnetoresistance~\cite{hamill2021two,Cho2022} that suggest a two-fold symmetric anisotropic superconducting gap and non-monotonic behavior of $T_{\mathrm{c}}$ with increasing disorder concentrations~\cite{Zhao2019,Rubio2020}.  

Motivated by these observations suggesting possible unconventional superconducting pairing and relevance of spin fluctuations in monolayer \nb{}, we study the leading superconducting instabilities in monolayer \nb{} within the assumption that the pairing is mediated solely by spin and charge fluctuations.
Previous works addressing Ising superconductivity from a repulsive pairing scenario have considered dominant pairing channels and symmetry arguments to obtain the superconducting gap symmetry~\cite{mockli2020ising, he2018magnetic, mockli2018robust,frigeri2004superconductivity,bulaevskii1976superconductivity,samokhin2008upper,ilic2017enhancement,mockli2019magnetic,Horhold_2023}. Here, we develop a general framework incorporating Ising SOC and the multi-orbital nature in the electronic structure, and calculate the pairing kernel within a full-fledged multi-orbital spin- and charge-fluctuation formalism \cite{berk1966effect,scalapino1999superconductivity,Romer2015,Kreisel2017,Romer2019,Kreisel2022}. We solve the associated linearized gap equation to obtain the leading superconducting order parameter and study the variation in solutions both as a function of SOC and Coulomb interaction strength. This leads to interesting regimes for the mixing between even- and odd-parity superconducting solutions. To identify the effect of the solution on measurable quantities, we next study the effect of an in-plane and out-of-plane magnetic field on $T_{\mathrm{c}}$ by solving the corresponding gap equations. Our work paves the way for a deeper understanding of the consequences of unconventional pairing in monolayer TMDs, and offers valuable insight into the interplay between Ising superconductivity with allowed singlet-triplet mixing and unique magnetic field dependence.\\
\noindent
This paper is organized as follows: In Sec.~\ref{sec:level2}, we provide a description of the applied model Hamiltonian and our calculations of the relevant susceptibilities. In Sec.~\ref{sec:level3}, we derive the pairing kernel and solve the associated gap equation to infer the preferred superconducting states for monolayer Ising superconductors. In Sec.~\ref{sec:level4} we discuss the effects of out-of-plane and in-plane magnetic fields. Finally, in Sec.~\ref{sec:level5}, we conclude and give a summary of our work. The associated Supplementary Material (SM) contains additional technical details.

\section{\label{sec:level2}Model and Methodology}

\subsection{Hamiltonian}
The lattice structure of monolayer \nb{} consists of a hexagonal lattice of Nb atoms sandwiched between Se atoms above and below the Nb plane.
The arrangement of the Se atoms leads to broken inversion symmetry, but retains the $\mathcal{M}_{xy}$ mirror plane. 
As we reduce the number of layers, the point group symmetry of the crystal reduces from a D$_{6h}$ symmetry (with inversion) relevant to the bulk material, to a D$_{3h}$ symmetry (without inversion) in the monolayer. 
In the presence of Ising SOC, the eigenstates of $S_z$ remain good spin quantum numbers, and
the SOC preferentially orients the spins to be along the $z$-axis i.e., out of the Nb plane. 
In the following, we utilize a multi-orbital tight-binding Hamiltonian to describe the low-energy electronic structure of monolayer \nb{}
\begin{eqnarray}
  H_{\rm TB}=\sum_{\vk l_1 l_2 \sigma} h_{l_1 l_2 \sigma}(\vk)c^{\dagger}_{\vk l_1 \sigma}c_{\vk l_2 \sigma}.\label{eq_tb}
\end{eqnarray}
Here, $c^{\dagger}_{\vk l \sigma}$ creates an electron with spin-$\sigma$, orbital $l$, and wavevector $\vk$. The Hamiltonian matrix with components $ h_{l_1 l_2 \sigma}(\vk)$ from Eq.~\eqref{eq_tb} can be expressed as a $6\times 6$ block diagonal matrix due to the absence of spin-flip terms in the SOC. The above Hamiltonian can be represented in the band basis by performing a unitary transformation
\begin{align}
 c_{\vk,\tilde{l}}=\sum_{\alpha}u^{\alpha}_{\tilde{l}}(\vk)a_{\vk \alpha \sigma},
\label{eq_unitary}
\end{align}
where $u^{\alpha}_{\tilde{l}}(\vk)$ are components of the eigenvector corresponding to band $\alpha$ and we have used a compact notation $[\tilde{l}:=(l,\sigma)]$. In Fig.~\ref{fig_ising_band_fs}(b) we show the bands crossing the Fermi level together with the orbital content, highlighting that the Fermi surface exhibits regions where each of the three orbitals dominates.
The model has been extracted from a DFT calculation using the full-potential local-orbital (FPLO) code \cite{Koepernik1999}, version 18.00-52 where a monolayer \nb{} was set up with the lattice constants $a=b=3.44\mathrm{\AA{}}$ and $c=12.48\mathrm{\AA{}}$ to simulate a vacuum. The space group is \# 187, i.e. P-6m2 with the Nb at the origin and the Se atom at $(1/3,-1/3,0.1322)$. We employed the nonrelativistic setting for the band structure without SOC and the fully relativistic setting to obtain the effects of Ising SOC. The Wannier projection was performed onto the three orbitals $d_{z^2}$, $d_{x^2-y^2}$, $d_{xy}$ with energy window of $[-0.4,3.5]\,\mathrm{eV}$.

We find that a minimal model consisting of three Nb orbitals $(d_{z^2},d_{x^2-y^2},d_{xy})$ reproduces well the DFT-generated electronic structure. 
The low-energy electronic structure yields a single electronic band forming the Fermi surface that splits into two spin momentum locked bands in the presence of Ising SOC of strength $\delta_{\vk}^{max}\sim 100$meV, in agreement with previous studies \cite{sticlet2019topological,bawden2016spin}. Note that the SOC is not used as a tuning parameter in our study, but the two electronic structures considered in this work: with SOC and without SOC are both derived from DFT calculations as described above. (More details on the electronic structure are presented in Sec-I of the SM). Similar realistic low energy tight binding models have been employed previously to study effect of spin orbit coupling on the electronic structure of various di-chalcogenide materials. These include a three band model (6 band including SOC) describing MoS$_{2}$, MoSe$_{2}$ and  MoTe$_{2}$ \cite{liu2013three}, a two band model (4 band when including magnetic field)\cite{falson2020type} describing type 2 Ising spin orbit coupling in Stanene, a five band d-p model (10 band when including SOC)\cite{Ridolfi_2015} describing MoS$_{2}$,
a four band model (8 band including SOC) describing  1T' -WTe$_{2}$\cite{lau2019influence} and
an eleven band tight-binding Hamiltonian (22 band including SOC)\cite{fang2015ab} describing WS$_2$, WSe$_2$ .

As shown in Fig.~\ref{fig_ising_band_fs}(a), the two low-energy bands forming the Fermi surface can be described by eigenstates $|\vk,\uparrow\rangle$,
and another band that enforces Kramers degeneracy with eigenstate $|-\vk,\downarrow\rangle$.
Both bands contain contributions from all three $d$-orbitals, and each band is described additionally by a particular spin eigenstate due to spin momentum locking. As seen in Fig.~\ref{fig_ising_band_fs}(b), the $d_{z^2}$ orbital dominates the orbital content of the $\Gamma$-centered Fermi pockets, whereas the $K$ and $K'$ centered pockets are dominated by the $d_{x^2-y^2}$ and $d_{xy}$ orbitals. In Fig.~\ref{fig_ising_band_fs}(c) we display the spin-resolved Fermi surface for monolayer \nb{} in the presence of Ising SOC. The presence of a horizontal mirror plane in the $xz$ plane ensures that the SOC and the band splitting vanishes along the $\Gamma-$M directions.

The bare electronic repulsion is given by the usual on-site Hubbard-Hund interaction, parametrized by $U$ and $J$. In a compact notation restricted to the Cooper channel, this interaction takes the form of (details given in SM)
\begin{eqnarray}
H_{\rm int}=\frac{1}{2} \sum_{\vk,\vk',\tilde{l}} [V]^{\tilde{l}_1,\tilde{l}_2}_{\tilde{l}_3,\tilde{l}_4}(\vk,\vk') \, c^{\dagger}_{\vk,\tilde{l}_1} \,c^{\dagger}_{-\vk,\tilde{l}_3} \, c_{-\vk',\tilde{l}_2}\, c_{\vk',\tilde{l}_4}.
\label{eqn:int_hamil}
\end{eqnarray}
Thus, the total Hamiltonian applied in the following analysis is given by
 \begin{eqnarray}
     H=H_{\rm TB}+H_{\rm int}\,.\label{eq_ham}
 \end{eqnarray}

\subsection{Calculation of Susceptibility}

Computation of the spin and charge susceptibilities allows to extract the structure of the dominant spin and charge fluctuations, which in turn dictate the superconducting pairing within the present approach. The multi-orbital non-interacting bare susceptibility tensor at momentum ${\vq}$ is defined as
\begin{align}
 \chi_{\tilde{l}_3 \tilde{l}_4 }^{\tilde{l}_1 \tilde{l}_2 }(\vq,\tau)\!=\!\!\frac{1}{N}\sum_{\vk,\vk'} & \langle T_{\tau} \, c^{\dagger}_{\vk,\tilde{l}_1}(\tau)c^{}_{\vk+\vq,\tilde{l}_2}(\tau)
 c^{\dagger}_{\vk',\tilde{l}_3}(0)  c_{\vk'-\vq,\tilde{l}_4}(0) \rangle_{0}.
\end{align}
Explicitly, this is given by
\begin{align}
\chi_{\tilde{l}_3 \tilde{l}_4  }^{\tilde{l}_1 \tilde{l}_2}(\vq,i \omega_{n})&=\frac{1}{N}\sum_{\vk,\alpha,\beta}
\frac{f(\epsilon_{\beta}(\vk+\vq))-f(\epsilon_{\alpha}(\vk))}{i\omega_{n} + \epsilon_{\alpha}(\vk)-\epsilon_{\beta}(\vk+\vq)}
\nonumber\\ &\times
u^{\alpha}_{\tilde{l}_4}(\vk)u^{\alpha *}_{\tilde{l}_1}(\vk)u^{\beta}_{\tilde{l}_2}(\vk+\vq)
u^{\beta *}_{\tilde{l}_3}(\vk+\vq)
.\label{eq_susceptibility}
\end{align}

The eigenenergies $\epsilon_{\alpha}(\vk)$ are calculated at momentum $\vk$ and belong to band $\alpha=(1,\ldots,6)$, and the function $f(x)=(1+e^{\beta x})^{-1}$ with $\beta=1/{k_BT}$ denotes the Fermi-Dirac distribution function at temperature $T$.

\begin{figure}
\centering
    \includegraphics[width=1.0\linewidth]{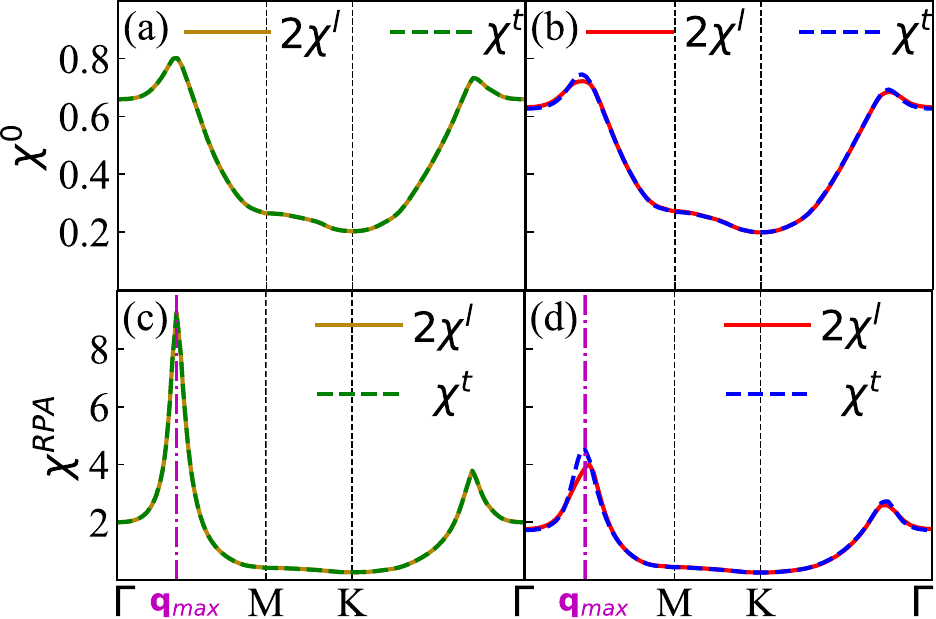}
    \caption{Spin susceptibility in units of 1/eV, at $\omega=0$ along the high-symmetry path $\Gamma$-$M$-$K$-$\Gamma$. (a) and (c) display the non-interacting and RPA
susceptibilities, respectively, calculated without SOC. (b) and (d) display these same quantities in the presence of Ising SOC. For both cases the RPA paramagnetic susceptibility peak occurs at a wave vector $\vq_{\mathrm{max}}\sim 0.4 $$\Gamma-$M. For the RPA susceptibility, we used $U=1.0$\;eV, $J=U/4$.
}
    \label{fig_xi}
 \end{figure}

From Eq.~\eqref{eq_susceptibility} we perform an analytical continuation $i\omega_n\rightarrow \omega +i\eta$ and evaluate the multi-orbital susceptibility by summing over a $k$-grid in the Brillouin zone (BZ) and take interactions into account via the random phase approximation (RPA), see SM for details. In Fig.~\ref{fig_xi}, we show the physical paramagnetic susceptibility extracted from the multi-orbital susceptibility. To extract the physical susceptibility from $\chi_{\tilde{l}_3 \tilde{l}_4 }^{\tilde{l}_1 \tilde{l}_2 }(\vq,i\omega_n)$ we consider the orbital components having $l_1=l_2$, and $l_3=l_4$ and sum over the orbitals. Then the physical susceptibility in the longitudinal ($\chi^{l}$) and transverse ($\chi^t$) channels can be expressed as
\begin{eqnarray}
 \chi^{l}(\vq,\omega)&=&  \frac{1}{4} (\chi_{\uparrow \uparrow }^{\uparrow \uparrow }-\chi_{\downarrow \downarrow }^{\uparrow \uparrow }
 -\chi_{\uparrow \uparrow }^{\downarrow \downarrow }+\chi_{\downarrow \downarrow }^{\downarrow \downarrow }), \\
 \chi^t(\vq,\omega)&=& \frac{1}{2} (\chi_{\downarrow \uparrow }^{\uparrow \downarrow }+\chi_{\uparrow \downarrow }^{\downarrow \uparrow }),
\end{eqnarray}
where, for ease of notation, we have suppressed the orbital indices that are being summed over.

The paramagnetic susceptibility peaks at a wave vector $\vq_{\mathrm{max}}\sim 0.4 $$\Gamma-$M, which agrees with previous calculations from a DFT-derived band structure \cite{kim2017quasiparticle,das2021quantitative,costa2022ising}.
Experimentally, a 3Q CDW order is observed in monolayer \nb{} which corresponds to a wave vector of $\vq_{\mathrm{CDW}}\sim 2/3\Gamma-$M
along the three symmetry-related directions \cite{zhang2022visualization}, implying that a pure nesting-driven scenario cannot explain the formation of CDW in this material.
Therefore, it is likely that additional effects, such as electron-phonon interactions, need to be accounted for in order to understand the CDW phase in monolayer \nb{}. A similar electron-phonon interaction effect has been used to explain the formation of 3Q CDW state in bulk bilayer 2H$-$\nb{} \cite{Johannes2006,Weber2011,flicker2016charge}.

In the presence of Ising SOC, the susceptibility is calculated in a basis where the Ising SOC induced spin momentum locking and corresponding spin splitting of electronic bands is taken into account. In Fig.~\ref{fig_xi}(c) and Fig.~\ref{fig_xi}(d), we show the non-interacting and RPA susceptibilities in the presence of Ising SOC, respectively.
The non-interacting susceptibility is slightly reduced upon the inclusion of Ising SOC, although the maximum peak remains at $\vq_{\mathrm{max}}$.
As expected, the SOC leads to a breaking of the degeneracy between the transverse and longitudinal channels. 
It turns out that in the presence of Ising SOC, the transverse susceptibility is enhanced compared to the longitudinal susceptibility, an effect that is much stronger in the interacting RPA susceptibilities. In the following section we show that such enhancement causes the interchange in the role of dominant even-parity superconducting instability and subdominant odd-parity instability. As the enhancement increases as a function of Coulomb interaction $U$, the odd-parity solution becomes the favorable state.

\section{\label{sec:level3}Leading Superconducting Instabilities}
Within a spin fluctuation-mediated pairing scenario, the effective superconducting pairing kernel has to incorporate the mixing of spin and orbital indices induced by the SOC~\cite{Romer2019}. The pairing kernel can be expressed as
\begin{eqnarray}
  [V(\vk,\vk')]^{\tilde{l}_1,\tilde{l}_2}_{\tilde{l}_3,\tilde{l}_4}&=&[U]^{\tilde{l}_1,\tilde{l}_2}_{\tilde{l}_3,\tilde{l}_4}
  -[\frac{U\,\chi_{0}\,U}{1-\chi_{0}\,U}]^{\tilde{l}_1,\tilde{l}_4}_{\tilde{l}_3,\tilde{l}_2}(\vk-\vk') \nonumber \\
  &+&[\frac{U\,\chi_{0}\,U}{1-\chi_{0}\,U}]^{\tilde{l}_1,\tilde{l}_2}_{\tilde{l}_3,\tilde{l}_4}(\vk+\vk').
\end{eqnarray}
\begin{figure}
    \includegraphics[width=1.0\linewidth]{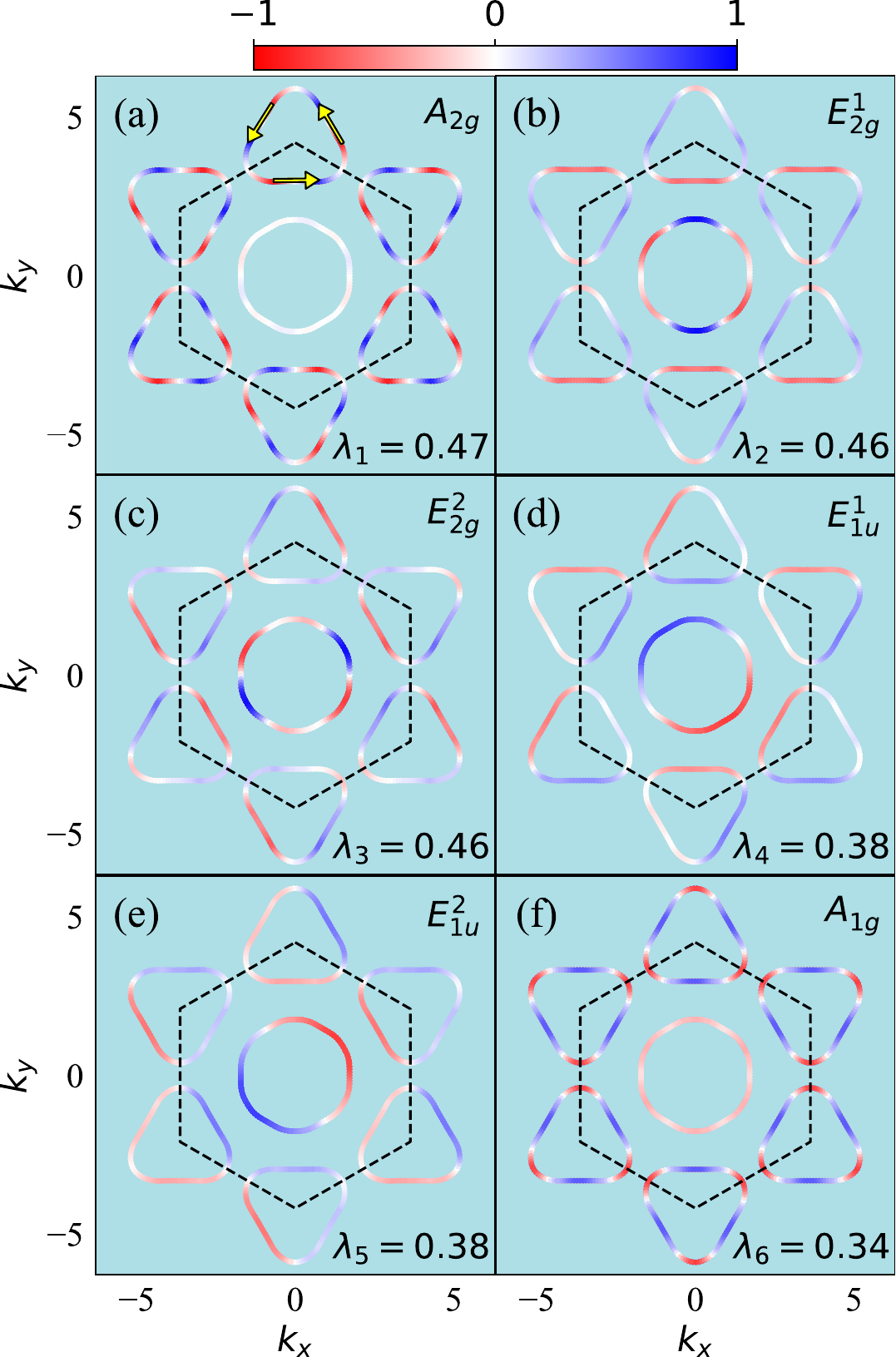}     \caption{Solutions of the linearized gap equation $\Delta(\vk_F)$
    plotted on the Fermi surface for the six leading superconducting pairing gaps with corresponding eigenvalue $\lambda$. The non-interacting Hamiltonian used here ignores the Ising SOC. Additional parameters: $U = 1.0$\;eV, $J = U/4$.    }
    \label{fig_with_out_ising_gap}
\end{figure}
The interaction matrix $U$ and the individual contributions to the pairing kernel are provided in the SM.  The superconducting gap function is defined as
\begin{small}
\begin{align}
 \Delta^{\tilde{m}_{1},\tilde{m}_2}(\vk)=-\sum_{\vk',\tilde{m}_{3},\tilde{m}_{4}}V^{\tilde{m}_2,\tilde{m}_1}_{\tilde{m}_3,\tilde{m}_4}(\vk,\vk')\langle a_{\vk,\tilde{m}_3}(\tau)a_{-\vk',\tilde{m}_{4}}(\tau) 
 \rangle,
 \end{align}
\end{small}
where $V_{\tilde{m}_3,\tilde{m}_4}^{\tilde{m}_1,\tilde{m}_2}(\vk,\vk')$ is the pairing interaction projected on the Fermi surface. The indices $[\tilde{m}_i:=(m_i,\sigma_i)]$ represent the band indices and the spin degrees of freedom, respectively; $m_i$ takes the values $(1,2,3)$ and $\sigma_i$ takes the values $(1,2)$.

The linearized gap equation in the presence of Ising SOC can be derived from the Gor'kov Green's functions (see SM Sec-III). For a two-band system with bands relevant to monolayer \nb, the gap equation can be expressed as
\begin{align}
\Delta^{\tilde{m},\tilde{m}'}_{}(\vk)=-\tilde{I}&\sum_{\substack{ \vk' }}\left[ V^{\tilde{m}',\tilde{m}}_{\tilde{m},\tilde{m}'}(\vk,\vk')
\Delta^{\tilde{m},\tilde{m}'}_{}(\vk') \right.\nonumber\\
 & \left.+   V^{\tilde{m}',\tilde{m}}_{\tilde{m}',\tilde{m}}(\vk,\vk') 
 \Delta^{\tilde{m}',\tilde{m}}_{}(\vk')\right],
 \label{Eq_gap_master}
\end{align}
where $\tilde{I} = \text{ln}\left( \frac{2e^{\gamma}\hbar\omega_{\mathrm{c}}}{\pi k_{B} T_{\mathrm{c}} }\right)$. Here the gap equation is restricted to the opposite spin pairing states with $\tilde{m}=(m,\sigma)$, and $\tilde{m}'=(m',\bar{\sigma})$.
A zero energy and a zero momentum gap function $\Delta^{\tilde{m},\tilde{m}'}(\vk)$ pairs opposite spin states, i.e. with different $\tilde m\neq \tilde m'$, but originating from the same $m$. We note that the projection onto the Fermi surface will only give contributions for bands crossing the Fermi level. For the NbSe$_2$ monolayer it turns out that only 2 of the 6 bands cross the Fermi level (see Fig. S1 in the supplementary materials) and we therefore only calculate the order parameters on these two bands.

\subsection{Gap solutions without Ising SOC}
\begin{figure}
    \includegraphics[width=1\columnwidth]{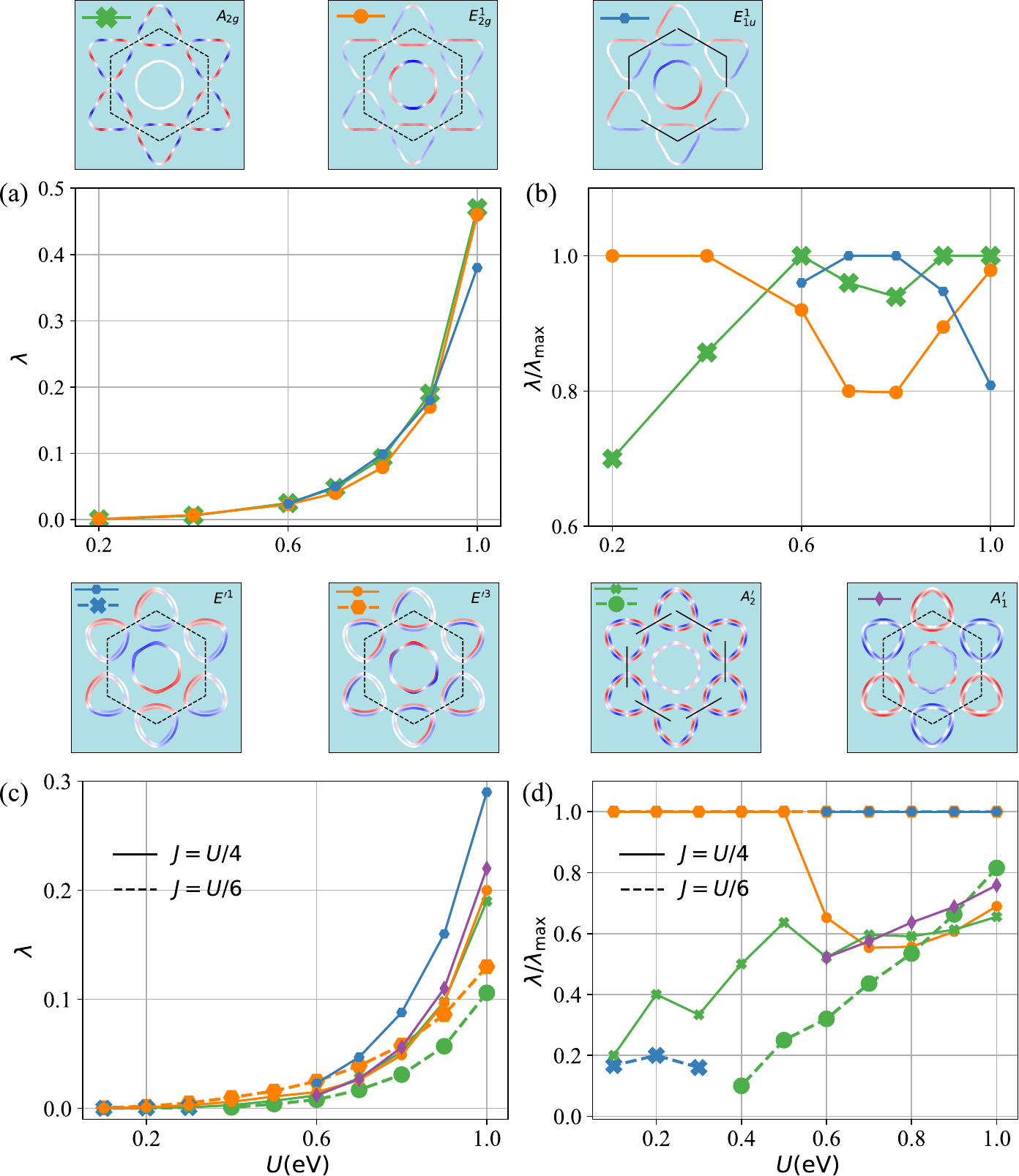}
    \caption{(a,b) Evolution of eigenvalues ($\lambda$, $\lambda/\lambda_{max}$), associated with a given superconducting gap function as a function of $U$ with $J = U/4$ in the absence of Ising SOC. (c,d) Evolution of the eigenvalue ($\lambda$, $\lambda/\lambda_{max}$), associated with a given superconducting gap function as a function of $U$ in the presence of Ising SOC. The solid lines correspond to $J = U/4$, and the dashed lines correspond to  $J = U/6$, respectively. The inset gap solutions displayed have the same color bar as Fig.~\ref{fig_with_out_ising_gap}. Note that for $U<0.6$\;eV, E$^{\prime 1}$ and  A$_1^{\prime}$ solutions have much smaller eigenvalue, and hence are not visible in the plot.
    }
     \label{fig:effect_of_j}
\end{figure}

Calculations of the superconducting instability from spin fluctuations in absence of SOC have been done in the literature for a number of multiband systems\cite{Graser2009,wang2013superconducting,kemper2010sensitivity,wu2015}. The gap function $\Delta(\bf k)$ of the leading instabilities and the corresponding eigenvalues can be obtained from the linarized gap equation
\begin{align}
\Delta^{m,m'}_{}(\vk)=-\tilde{I}&\sum_{\substack{ \vk' }} V^{m',m}_{m,m'}(\vk,\vk')
\Delta^{m,m'}_{}(\vk').
\end{align}
$T_\mathrm c$ is related to the eigenvalue $\lambda$ via $\exp{\left(-\frac{1}{\lambda}\right)}$. We apply this method to our Hamiltonian, Eq.~(\ref{eq_ham}) and show in Fig~\ref{fig_with_out_ising_gap}
the solutions of the linearized gap equation for $U=1$\,eV and in the absence of SOC where the underlying symmetry of the gap functions belongs to the irreducible representations (Irrep) of D$_{6h}$ point group. We show the dominant superconducting instability and the first 5 sub dominant instabilities. The ground state superconducting gap belongs to an Irrep of the even-parity A$_{2g}$ representation with gap nodes along the $\Gamma-$K directions. The dominance of the observed nodal pairing structure can be attributed to the nesting properties of the Fermi surface. As shown in Fig.~\ref{fig_xi}, the peak in the susceptibility occurs around 0.4$\Gamma$M. This momentum vector corresponds to the intra-pocket nesting vectors depicted by the yellow arrows in Fig.~\ref{fig_with_out_ising_gap}(a) (see also Sec-IV of SM for more details). We also find that the dominant A$_{2g}$ symmetry superconducting gap on the $\Gamma$-centered Fermi pocket is much smaller than the gap magnitude on the $K$-centered pockets. The subleading gap functions are a set of degenerate states that belong to the two-dimensional (2D) E$_{2g}$ Irrep (Fig.~\ref{fig_with_out_ising_gap}(b,c)) with basis functions ($f_{k_{x^2}-k_{y^2}}$,$f_{k_xk_y}$) followed by the odd-parity solutions belonging to the 2D E$_{1u}$ Irrep (Fig.~\ref{fig_with_out_ising_gap}(d,e)) with basis functions $(f_{k_x},f_{k_y})$. In fact, as seen from Fig.~\ref{fig:effect_of_j}(a,b), the gap functions belonging to these three states remain quite close in energy with variation in Coulomb interaction. In this figure the dimensionless quantity $\lambda/\lambda_{max}$ is proportional to the ratio of log of the corresponding transition temperatures. One can clearly see that for $U<0.6$\;eV the dominant instabilities favour the two dimensional E$_{2g}$ symmetry, whereas for $U\geq 0.6$\;eV solutions corresponding to odd parity E$_{1u}$ or even parity A$_{2g}$ symmetry solutions are favoured.

\begin{figure}
     \includegraphics[width=1.0\linewidth]{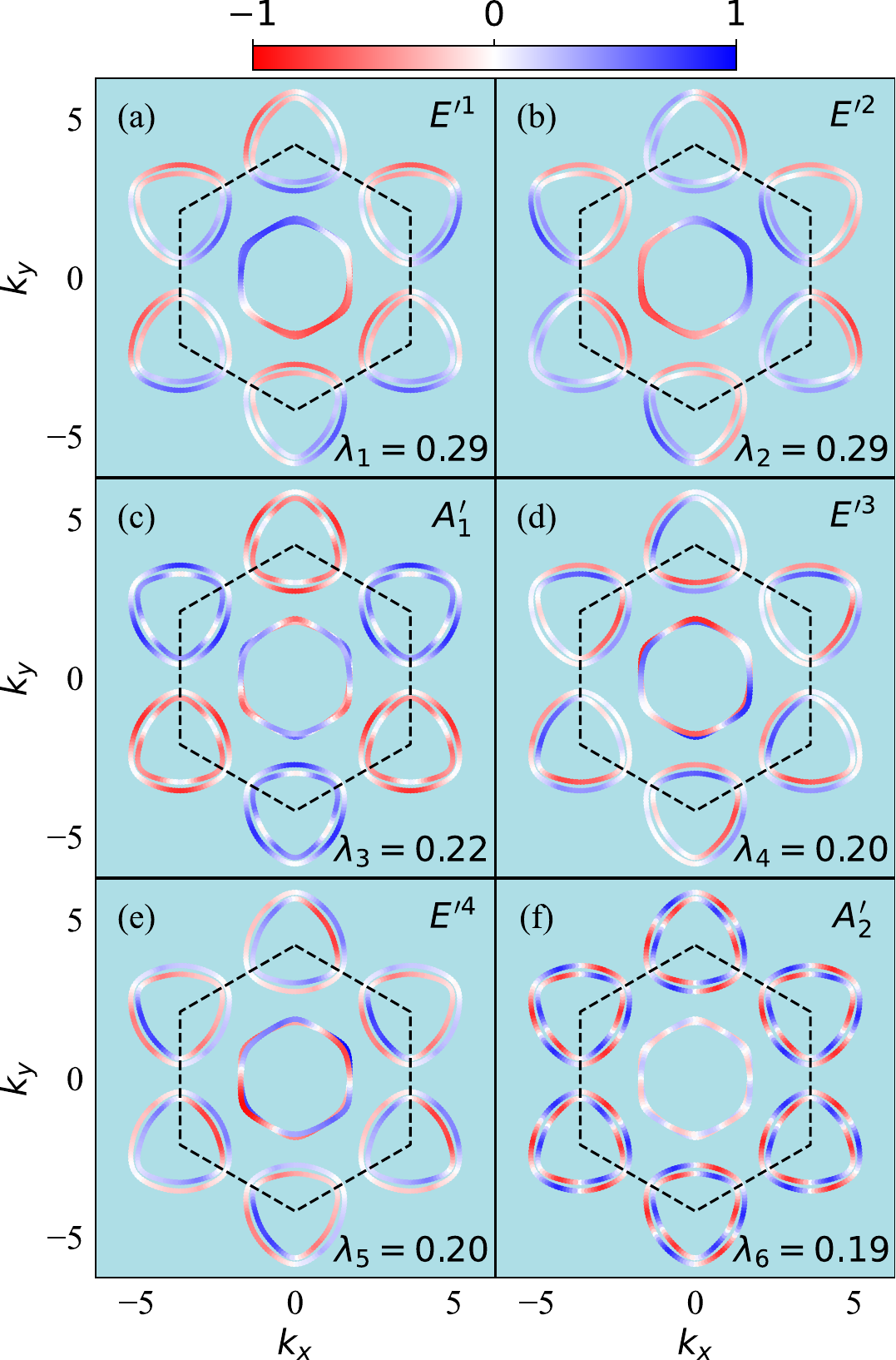}
    \caption{Superconducting gap functions plotted on Fermi surface in the presence of Ising SOC with $U = 1.0$\;eV, $J = U/4$. Here, $\lambda$ on the plots denote the eigenvalues of the corresponding solutions of the gap equation. Panels (a) and (b) display the degenerate ground state solution.}
    \label{fig_gap_ising}
\end{figure}

\subsection{Gap solutions with Ising SOC}
The gap equation, Eq.~\eqref{Eq_gap_master} contains spin-dependent pairing interaction from the transverse and longitudinal susceptibilites. Now, since the Fermi surface is spin-split, we need to set up the linearized gap equation with $\vk$ points on the two Fermi surfaces. This leads to a larger gap matrix equation and the corresponding gap function pairs electrons that formally belong to different bands (details in the SM).
In the following, we discuss our results for the pairing in the presence of Ising SOC. Previous studies considering a repulsive pairing interaction have proposed a leading superconducting instability in the $s+f$-wave channel \cite{Horhold_2023,aliabad2018proximity,mockli2020magnetic,mockli2019magnetic,Black-Schaffer_2014}. In Fig.~\ref{fig:effect_of_j}(b,d) we show the evolution of the various gap function symmetries with variation in Coulomb interaction. Here, the gap function solution belonging to the same irreducible representation E$^{'}$ has been shown separately with notation E$^{'^{1}}$  and E$^{'^{3}}$ since these two solutions show different admixture with triplet components. In particular, the solution E$^{'{^1}}$ has a stronger odd parity admixture compared to the E$^{'^{3}}$ solution.  In Fig.~\ref{fig:effect_of_j}(d) we present the ratio $\lambda/\lambda_{max}$ that is proportional to the ratio of log of transition temperature. It is apparent from the figure that the leading instability for $U<0.6$\;eV belongs to the E$^{'^{3}}$ Irrep. of D$_{3h}$ point group. However, the relative mixing of the triplet component is enhanced for larger Hunds coupling $J$. As can be seen from Fig. \ref{fig:effect_of_j}(d) the E$^{'^{1}}$(E$^{'^{3}}$) solution gives the dominant instability for $U>0.6$\;eV when Hunds coupling $J=U/4$ ($J=U/6$) respectively. In order to focus on the detailed momentum space structure of the gap function, in Fig.~\ref{fig_gap_ising} we present the solutions of the linearized gap equation in the presence of Ising SOC for $U=1$\,eV, and $J=U/4$.

\begin{figure}
    \centering
    \includegraphics[width=1.0\linewidth]{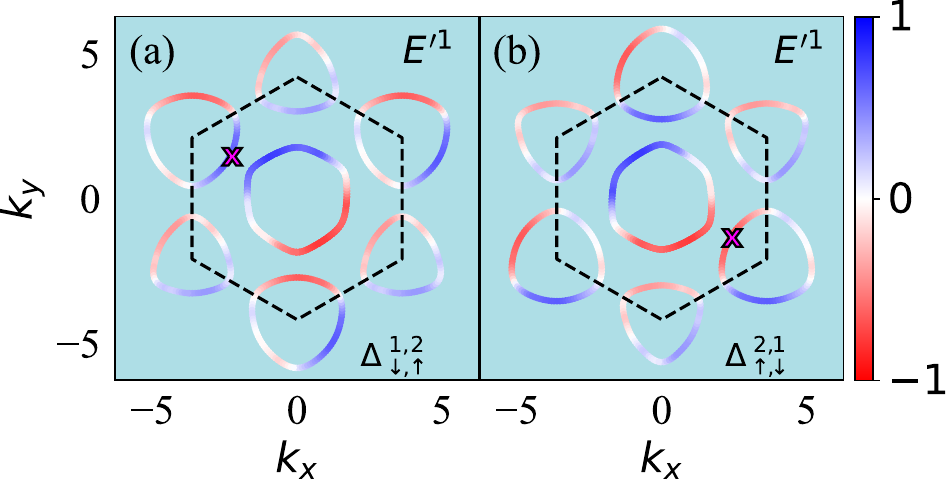}
    \caption{The ground state superconducting gap function displayed separately on the spin-up (a) and spin-down (b) Fermi surfaces. The solution on the Fermi pocket on the two bands are related by $\Delta^{1,2}_{\uparrow,\downarrow}(\vk)=-\Delta^{2,1}_{\downarrow,\uparrow}(-\vk)$. Here, $U = 1$, $J = U/4$ and $\lambda = 0.29$.}
    \label{fig_spin_resolved_gap}
\end{figure}

In the following, the notation for band indices $\tilde m=(1,2)$ correspond to bands with definite spin eigenstates $(\uparrow,\downarrow)$ respectively. From Fig.~\ref{fig_gap_ising} it is apparent that the ground state superconducting gap solution exhibits a degeneracy between the first two obtained eigenvalues (see Fig.~\ref{fig_gap_ising}(a,b)), and belongs to the 2D E$'$ Irrep of the D$_{3h}$ point group.  In the presence of broken inversion symmetry, this solution mixes  even- and odd-parity states corresponding to basis functions of E$_{2g}$ and E$_{1u}$ symmetry of the D$_{6h}$ point group, although as discussed below the odd-parity contribution dominates the mixed state for stronger Coulomb interactions. This result deviates from the corresponding solution in absence of SOC where the ground state superconducting solution is an even-parity A$_{2g}$ ($i$ wave) state whereas the A$_2'$ symmetry superconducting state (A$_{2g}$+B$_{2u}$ or $i+f$) is a subdominant order in the presence of SOC (see Fig.~\ref{fig_gap_ising}(f)). We also find that the superconducting gap on the $\Gamma$-centered pocket is relatively larger in the presence of Ising SOC and becomes comparable to the gap on the $K$-centered pockets for the dominant E$_2$ as well as the subdominant A$_1$ representations. These results are supported by the corresponding nesting vectors for the spin-split Fermi surface. The dominant nesting in the presence of Ising SOC connects additional inter-pocket regions between the $K$ and $K'$ points as well as regions between the $\Gamma$-pocket and the $K$-pockets, which is the likely reason for the enhanced superconducting gap on the $\Gamma$ pocket in the presence of SOC. (see Sec-IV of SM for further details). Further, considering a model where we only include the Ising SOC in the electronic band structure and not in the superconducting pairing kernel, we find that the ground state superconducting solution still exhibits dominant odd-parity state mixture. This implies that the spin-split band structure due to presence of the Ising SOC preferentially supports an odd-parity superconducting state.

The solution on the Fermi pocket on the two bands are related by $\Delta^{1,2}_{\uparrow,\downarrow}(\vk)=-\Delta^{2,1}_{\downarrow,\uparrow}(-\vk)$, see Fig.~\ref{fig_spin_resolved_gap}(a,b), which is required by the exclusion principle. In order to identify the superconducting gap symmetry and corresponding Irrep from the gap structure, consider a $180^{\circ}$ rotation operation $\sigma_v$ about an in-plane axis slanted at an angle of $30^{\circ}$ from the $k_x$ direction that we call the $k_x'$ direction. If we express the rotated coordinate system in a $(k_x',k_y')$ basis, then $k_x'=(\sqrt{3}k_x+k_y)/2$, and $k_y'=(-k_x+\sqrt{3}k_y)/2$. Therefore, since from Fig.~\ref{fig_spin_resolved_gap}(a,b) $\sigma_v\Delta(k_x',k_y')=\Delta(k_x',-k_y')=-\Delta(k_x',k_y')$, we have a node at $\Delta(k_x',0)$. This transformation of the gap function under a symmetry operation of the D$_{3h}$ point group combined with the presence of a degenerate superconducting state in the solution of the linearized gap equation, ensures that the corresponding gap function belongs to the E$'$ Irrep. In fact, the gap function can be expressed as $\Delta (\vk)\sim af_{k_x'k_y'}+bf_{k_y'}$, where $a$, and $b$ are constant numbers. In the original $(k_x,k_y)$ basis, the other components of the 2D Irrep will of course get mixed, but determining the specific mixing that minimizes the ground state energy would require the theory to be extended beyond a linearized gap equation.

\begin{figure}
    \centering
    \includegraphics[width=1.0\linewidth]{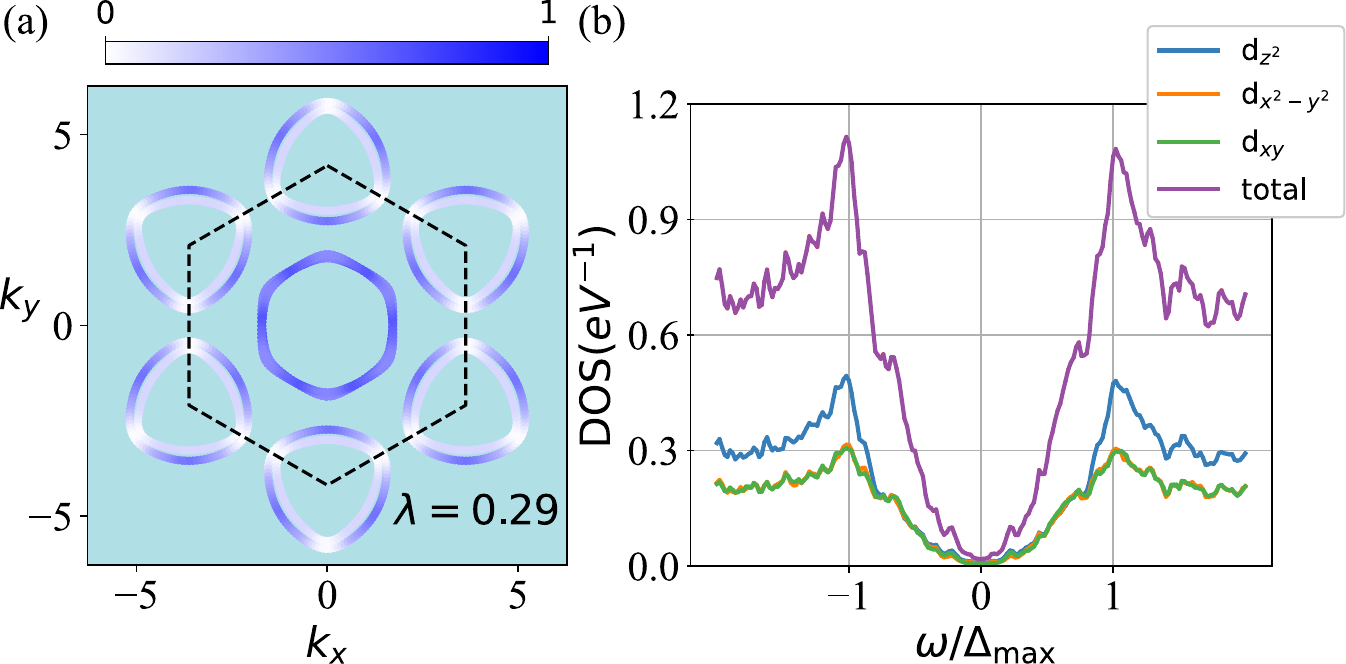}
    \caption{(a) $|\Delta_1(\vk)+i\Delta_2(\vk)|$ plotted on the Fermi surface pockets. Here $\Delta_1(\vk)$, and $\Delta_2(\vk)$ are the two degenerate leading superconducting gap functions shown in Fig.~\ref{fig_gap_ising}(a,b) for the case with Ising SOC.  (b) Orbital-resolved DOS corresponding to the solution $\Delta_1(\vk)+i\Delta_2(\vk)$.
    }
    \label{fig_d6h}
\end{figure}

It is instructive to consider the structure of $|\Delta_1(\vk)+i\Delta_2(\vk)|$ where $\Delta_1(\vk)$, and $\Delta_2(\vk)$ are the degenerate superconducting ground state solutions represented in Fig.~\ref{fig_gap_ising}(a,b). This complex superposition is the expected $T=0$ state due to its minimization of the condensation energy, as compared to the individual solutions. As shown in Fig.~\ref{fig_d6h}(a), the magnitude of this $(1,i)$ state belongs to an A$'$ representation of the D$_{3h}$ point group symmetry as expected from the decomposition of the reducible representation $\Gamma_{E'}\times \Gamma_{E'}^*$. The anisotropy of the gap structure leads to a multigap feature and a density of states (DOS) that displays a U-shaped form at low energies (see Fig.~\ref{fig_d6h}(b))\cite{kuzmanovic2022tunneling}. Although, these multigap features in the DOS would presumably be further influenced by the presence of an underlying CDW ordering, it is interesting to note that similar multigap features and a low-energy U-shaped DOS have been observed in low-temperature STM measurements on ultra-thin films of \nb \cite{dvir2018spectroscopy}. Additionally, the presence of closely lying sub dominant orders is a feature that can be obtained in calculation of superconducting state in multi orbital systems. Two prominent examples would be the case of iron based superconductors \cite{Graser2009} and Strontium Ruthenate\cite{Romer2019,Clepkens2021}. An interesting consequence of closely lying sub dominant order can be the formation of accidental degeneracy scenarios or observation of sub dominant superconducting gap structure symmetries in experiments performed with symmetry breaking external perturbations.

\subsection{Singlet-Triplet mixing}
\begin{figure}
    \centering
    \includegraphics[width=1.0\linewidth]{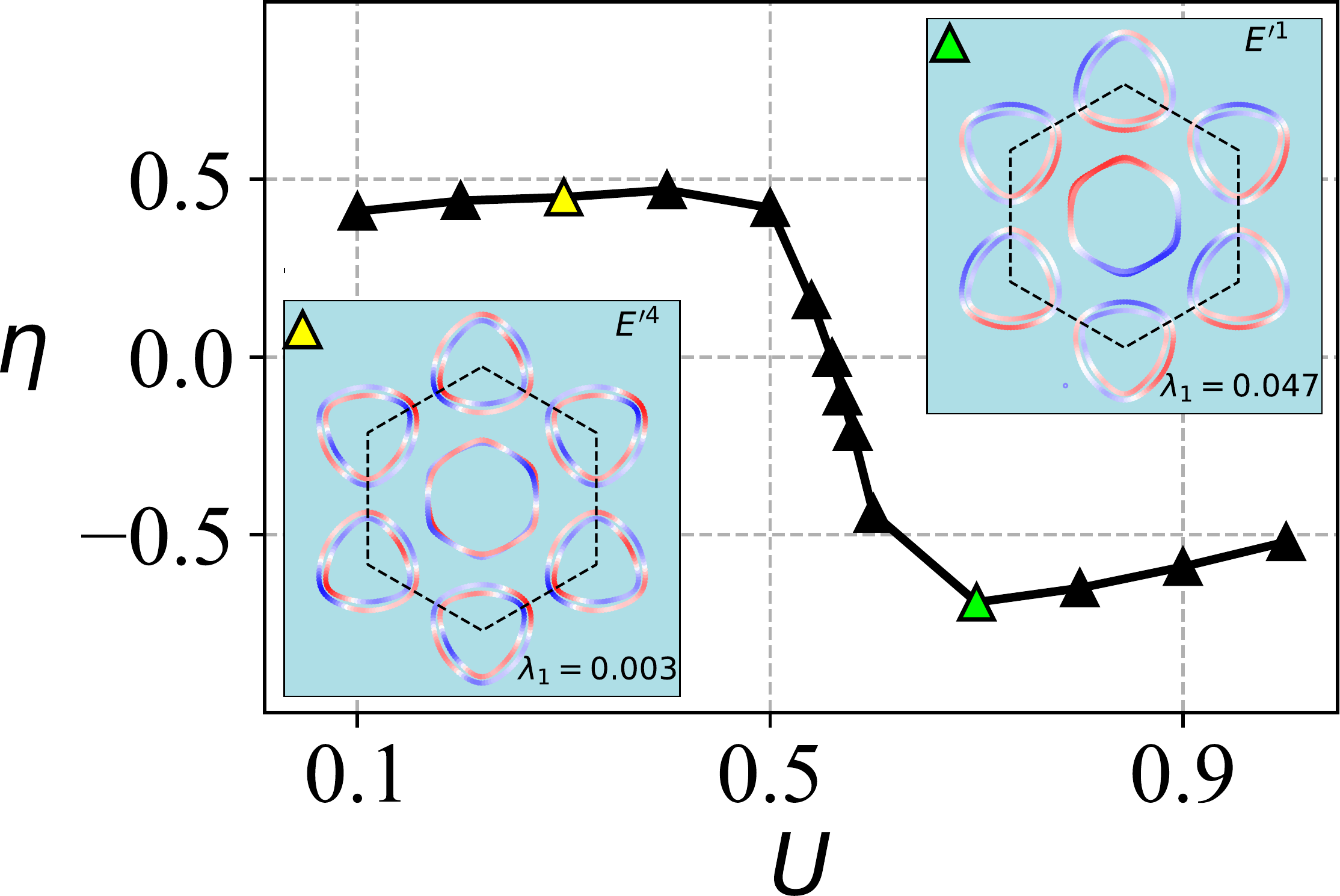}
    \caption{
    Superconducting singlet-triplet mixing ratio $\eta$ as a function of the Coulomb interaction $U$ with $J= U/4$. 
    $ \eta= (+1)(-1) $ corresponds to the pure even(odd) parity solutions. 
   The gap solutions displayed in the two insets have the same color bar as Fig.~\ref{fig_with_out_ising_gap}.
   }
    \label{fig:mixing_ratio}
\end{figure}

We can approximately quantify the singlet-triplet mixing by considering the superconducting gaps $\tilde{\Delta}^{1,2}_{\uparrow, \downarrow}(\vk)=\Delta^{1,2}_{\uparrow, \downarrow}(\vk+\phi_{\vk}/2)$ on the spin-up Fermi surface and $\tilde{\Delta}^{2,1}_{\downarrow \uparrow}(\vk)=\Delta^{2,1}_{\uparrow, \downarrow}(\vk-\phi_{\vk}/2)$ on the spin-down Fermi surface. At every Fermi wave vector $\vk$ corresponding to the Fermi surface without SOC, we define a $\phi_{\vk}$  representing half of the corresponding band splitting when SOC is introduced.
Then $\vk \pm \phi_{\vk}$ are approximately (the splitting, in general, will not be symmetrical) the Fermi wave vectors for the two spin-split bands. We then define the even- and odd-parity gaps as $\Delta_{o/e}(\vk)=W_{1}(\vk)\tilde{\Delta}^{1,2}_{\uparrow \downarrow}(\vk) \pm W_{2}(\vk)\tilde{\Delta}^{2,1}_{\downarrow \uparrow}(\vk)$ with weighting factor $W_{\tilde{m}}(\vk)=\frac{dl^{\tilde{m}}_{\vk}}{|\nabla\epsilon_{\tilde{m}}(\vk)|}/\sum_{\vk}\frac{dl^{\tilde{m}}_{\vk}}{|\nabla\epsilon_{\tilde{m}}(\vk)|}$,
and the quantity $\eta=\frac{1}{N_{\vk}}\sum_{\vk}\frac{|\Delta_e(\vk)|-|\Delta_o(\vk)|}{|\Delta_e(\vk)|+|\Delta_o(\vk)|}$, as a measure of the singlet-triplet mixing ratio, $N_{\vk}$ being the normalization factor. In Fig.~\ref{fig:mixing_ratio}, the variation of the mixing ratio $\eta$ of the ground state gap function reveals that the odd-parity superconducting state dominates the mixed state for larger Coulomb interactions. 

The observation of superconducting collective modes in tunneling experiments have been interpreted either as a Leggett mode between an s-wave state and a proximate f-wave channel~\cite{wan2022observation} or as a fluctuation of the superconducting order parameter phase between the $K$ and $\Gamma$ pocket induced by a strong suppression in gap over the $\Gamma$ pocket \cite{das2023electron}. Although our result does not find either a dominant $s+f$ gap structure, or strong anisotropy of the gap magnitude between the $K$ and $\Gamma$ pockets, we do find that within a spin-fluctuation-mediated pairing mechanism a significant mixing of even- and odd-parity superconducting states occur, and this could lead to a Leggett mode\cite{semenov2024pairing} between an E$_{1g}$ and a proximate E$_{1u}$ odd-parity channel.

\section{\label{sec:level4}Magnetic Field Effect}
The presence of Ising SOC and the resulting mixing of even- and odd-parity gap functions in TMD superconductors lead to a non-trivial dependence of the superconducting gap function on a magnetic field. The general Hamiltonian for the Zeeman term with arbitrary field direction can be written as
\begin{eqnarray}    H_{\mathrm{Zeeman}}=g\mu_{B}\sum_{l,\vk,\sigma,\sigma'}\left[\boldsymbol{\sigma}\cdot \mathbf{B}\right]_{\sigma\sigma'} c^{\dagger}_{\vk l \sigma}c_{\vk l \sigma'}.
\end{eqnarray}
Here, $\boldsymbol{\sigma} \equiv (\sigma_{x},\sigma_{y},\sigma_{z})$ denotes a vector of Pauli matrices and $\mathbf{B} \equiv(B_{x},B_{y},B_{z})$ originates from an applied magnetic field. In the following, the linearized superconducting gap equation is solved in the presence of a magnetic field. The model assumes that the external magnetic field primarily enters the Hamiltonian as a Zeeman field contribution, but does not modify the spin fluctuation mediated pairing interaction (spin fluctuations are considered in zero magnetic field). This also implies that the magnetic field effect is computed for a homogeneous superconducting state.

\subsection{Out-of-plane magnetic field}
\begin{figure}
    \centering
    \includegraphics[width=0.85\linewidth]{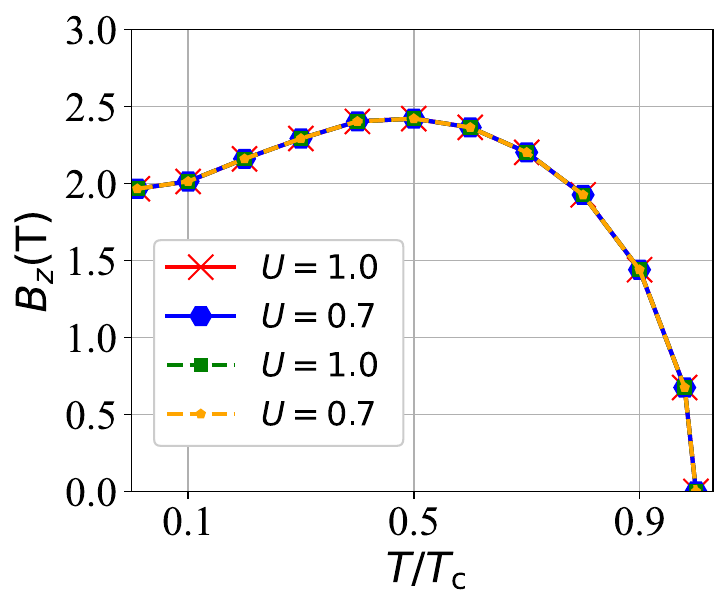}
    \caption{Effect of an out-of-plane magnetic field on $T/T_{c}$ for various $U$. Solid lines are obtained the presence of Ising SOC, and dashed lines are obtained in the absence of Ising SOC, respectively. (Solutions  may change at lower temperatures.)}
    \label{fig:out_of_plane_magnetic_field}
\end{figure}

We first study the effect of a longitudinal magnetic field on superconducting monolayer NbSe$_2$ ignoring any orbital effect and considering only the Pauli paramagnetic contribution within the superconducting state. The Zeeman field Hamiltonian is given by
\begin{eqnarray}
H_{B_z}=g\mu_B B_z \sum_{l,\vk}(n_{\vk l\uparrow}-n_{\vk l \downarrow}),
\end{eqnarray}
\noindent where $n_{\vk l \sigma}=c^{\dag}_{\vk l \sigma}c_{\vk l \sigma}$. Since the non-interacting Hamiltonian exhibits no spin-flip contributions, the above term preserves $S_z$ as a good quantum number. We consider the magnetic field as a perturbation and obtain the homogeneous solution of the linearized gap equation. This leads to a second-order transition curve similar to earlier studies~\cite{maki1964pauli}. The effect of out of plane magnetic field can be captured by the following equation (Detailed steps given in SM) \\
\begin{eqnarray}
  \ln\left(\frac{T}{T_c}\right)=-\Bigl\{ \mathrm{Re}\Bigl[\psi(\frac{1}{2}+i\frac{g\mu_{B}B_z}{2\pi K_{B}T})\Bigr] -\psi(\frac{1}{2})\Bigr\}
\end{eqnarray}
Here, $\psi(x)$ is the di-Gamma function and $T_{c}$ is the superconducting critical temperature in absence of magnetic field and is given by
\begin{eqnarray}
    T_{c}=\frac{2 \exp{(\gamma)}\hbar \omega_{c}}{\pi k_{\mathrm B}}\exp{\left(-\frac{1}{\lambda}\right)}
\end{eqnarray}
In Fig. 9 we depict the effect of an out-of-plane magnetic field on $T /T_{c}$ for various $U$. Solid lines are obtained in the presence of Ising SOC, and dashed lines are obtained in the absence of Ising SOC, respectively. We find all curves fall on top of each other, i.e. the even parity and the odd parity solutions are equally strongly affected by the field. In supplementary section this has been derived within Gorkov formalism in the section-III-(B). As shown in Fig.~\ref{fig:out_of_plane_magnetic_field}, we do not find any appreciable effect of the Ising SOC on the second-order phase transition curve in the case of out-of-plane magnetic fields.

\subsection{In-plane magnetic field}
For an in-plane magnetic field in a 2D material, the presence of a spin-flip term implies that $S_z$ is no longer a good quantum number. In this case, for a field in the $x$ direction, the Zeeman term contribution can be expressed as
\begin{eqnarray}
H_{B_x}=g\mu_B B_x\sum_{\vk,l} (c^{\dag}_{\vk l \uparrow}c_{\vk l \downarrow}+c^{\dag}_{\vk l\downarrow}c_{\vk l\uparrow}).
\end{eqnarray}
In the following, we consider the in-plane magnetic field as a perturbation and represent it in the band basis of the non-interacting Hamiltonian. This results in a Hamiltonian of the form
\begin{eqnarray}
H_{B_x}=\sum_{m m' \vk}\left(h_{\vk \uparrow \downarrow}^{m,m'}a_{\vk m \uparrow}^{\dag}a_{\vk m' \downarrow}+\mbox{H.c}\right).
\end{eqnarray}
Here, $h_{\vk \sigma \bar{\sigma}}^{m,m'}=g\mu_BB_x\sum_{\mu}u_{m \sigma}^{*\mu}(\vk)u_{m' \bar{\sigma}}^{\mu}(\vk)$ and we have used the transformation
in Eq.~(\ref{eq_unitary}) in zero magnetic field, $m$ represents the band index running from 1 to 3. Due to spin-spitting, we get 6 effective bands, 3 corresponding to spin-up and the other 3 corresponding to spin-down. The orbital index $\mu$ also runs over three orbital components ($d_{z^2}$, $d_{x^2-y^2}$, $d_{xy}$).
Since we ignore the effect of an in-plane magnetic field on the pairing kernel, our model does not consider the possibility of an equal-spin pairing gap induced by tilting of the spins from the out-of-plane direction induced by finite in-plane magnetic fields. However, as discussed in the SM, the presence of the in-plane Zeeman field leads to finite equal-spin superconducting correlations. Computationally, the resultant gap equation is numerically solved for each magnetic field value by fixing the superconducting transition temperature at zero field to $T_{\mathrm c_{}}=3$\;K. The linearized gap equation becomes
\begin{widetext}
 \begin{align}\label{fullgapeqn}
-\Delta^{\tilde{m},\tilde{m}'}_{}(\vk)=&\sum_{\substack{\vk' }}[ \left(  V^{\tilde{m}',\tilde{m}}_{\tilde{m},\tilde{m}'}(\vk,\vk') I^{}_{1}(\vk')
+
   V^{\tilde{m}',\tilde{m}}_{\tilde{m}',\tilde{m}}(\vk,\vk') I^{}_{2}(\vk')\right)\Delta^{\tilde{m},\tilde{m}'}_{}(\vk')\nonumber\\
 &+ \left( V^{\tilde{m}',\tilde{m}}_{\tilde{m}',\tilde{m}}(\vk,\vk')  I^{}_{1}(\vk')
  + V^{\tilde{m}',\tilde{m}}_{\tilde{m},\tilde{m}'}(\vk,\vk') I^{}_{2}(\vk')\right)\Delta^{\tilde{m}',\tilde{m}}_{}(\vk')],
\end{align}
\end{widetext}
where
\begin{subequations}
\begin{align}
    I_{1}&=- \text{ln}\left(\frac{T_{}}{T_{\mathrm{c_{}}}}\right)+\frac{1}{\lambda}\nonumber\\
    &- \frac{1}{2}\left(\frac{-h_{\vk}^{2}}{h_{\vk}^{2}+\delta_{\vk}^{2}}\right)
    \left\{\mathrm{Re}[\psi(\frac{1}{2}+i\frac{\sqrt{h_{\vk}^{2}+\delta_{\vk}^{2}}}{2\pi k_{B}T_{}})] -\psi(\frac{1}{2})\right\},\\
      I_{2}&=\frac{1}{2}\left(\frac{-h_{\vk}^{2}}{h_{\vk}^{2}+\delta_{\vk}^{2}}\right)\left\{\mathrm{Re}[\psi(\frac{1}{2}+i\frac{\sqrt{h_{\vk}^{2}+\delta_{\vk}^{2}}}{2\pi k_{B}T_{}})] -\psi(\frac{1}{2})\right\}.
\end{align}
\end{subequations}
\begin{figure}
    \centering
     \includegraphics[width=1.0\linewidth]{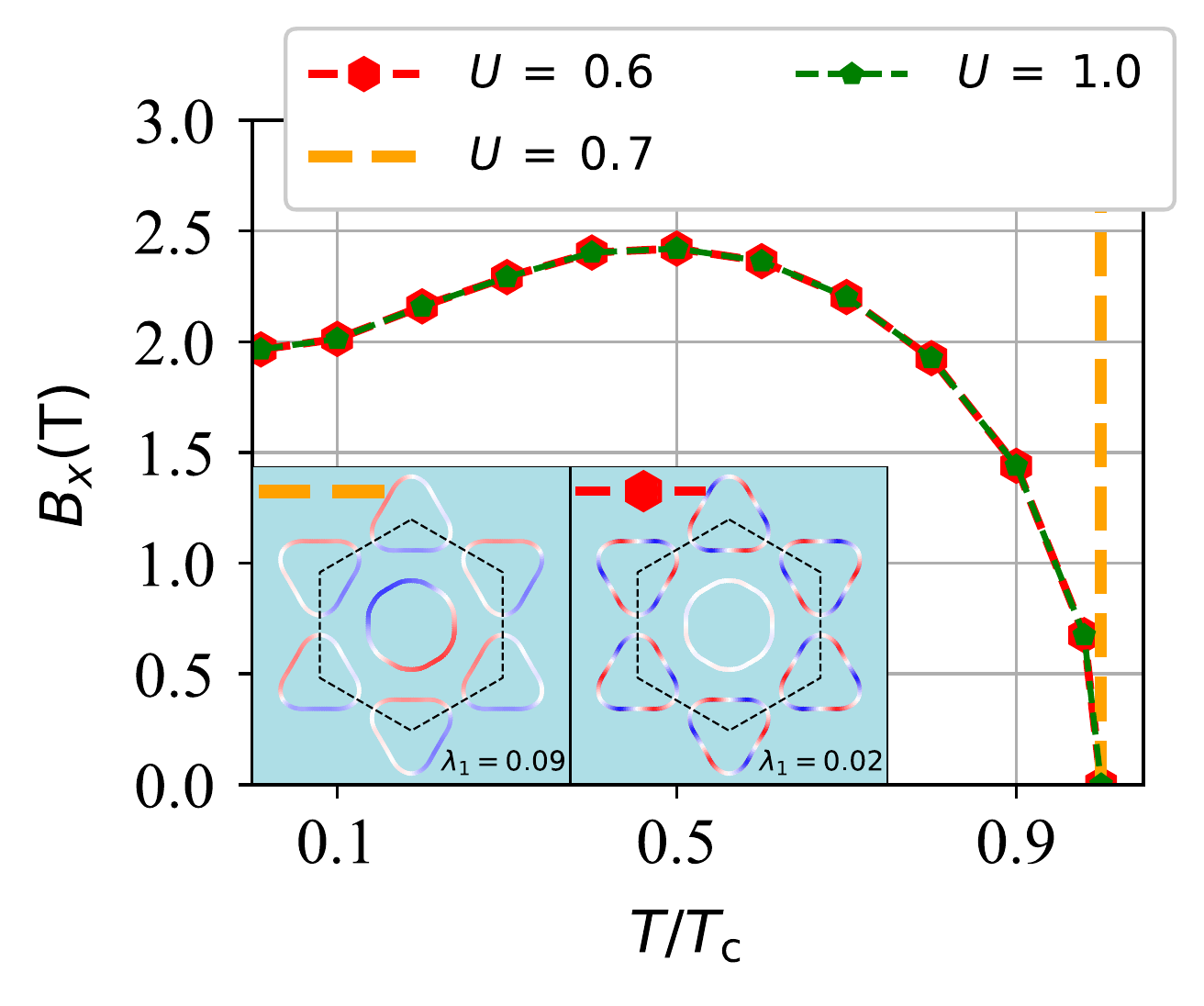}
    \caption{Effect of an in-plane magnetic field on $T/T_{\mathrm{c}}$ in the absence of SOC for various $U$. (Inset) Solutions corresponding to the highest eigenvalue for various $U$ in the absence of SOC.
   The inset gap solutions displayed have the same color bar as Fig.~\ref{fig_with_out_ising_gap}. All even parity (nominally singlet) solutions exhibit a strong sensitivity to the applied field. In contrast, odd parity (nominally triplet) solutions remain unaffected. There is the vertical line, i.e. in this approximation the critical field is ``infinite''. }
    \label{fig_in_plane_mag_with_out_ising}
\end{figure}
Here, $\psi$ denotes the digamma function and we define $2\delta_{\vk}$ as representing a momentum dependant energy gap between two SOC split bands near the Fermi level. Further detailed steps can be found in the SM. Let us first consider the solution of the gap equation in the absence of Ising SOC. In this case, we retain the classification of the solution to even- and odd-parity channels, and the spin index on the eigenvector components can be suppressed without loss of generality. Then the orthonormality of the eigenvectors allows us to write $h_{\vk \uparrow \downarrow}^{m,m'}=h_{x}$.  As can be seen from Fig.~\ref{fig_in_plane_mag_with_out_ising}, in the absence of SOC, the triplet pairing is unaffected by an in-plane magnetic field, whereas the singlet state gets suppressed~\cite{sigrist2009introduction}. This behavior agrees with previous studies of magnetic field effects in spin-singlet and spin-triplet superconductors \cite{maki1964pauli,mockli2020ising,mockli2020magnetic}. Note that the invariance of the transition temperature on the in-plane magnetic field for a triplet solution of monolayer NbSe$_2$ can be easily understood since the opposite-spin pairing triplet state implies that the $\vd(\vk)$~\cite{sigrist2009introduction} vector is out of the plane. Therefore, the transition temperature, which depends on $|\vd (\vk)\cdot \vB|$, is independent of the magnetic field.
\begin{figure}[]
    \centering
        \includegraphics[width=0.85\linewidth]{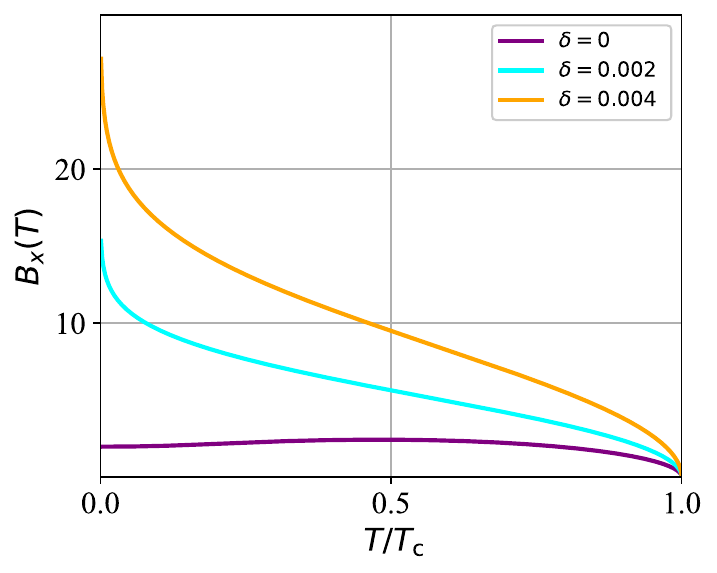}
    \caption{Effect of Ising SOC on $T/T_{\mathrm{c}}$ demonstrated in a simplified model, where we ignore the momentum-dependence of Ising SOC and consider an even-parity superconducting state obeying Eq.~(\ref{eqn_toy_model}).} 
     \label{fig:toy}
\end{figure}

In the presence of Ising SOC, the results become non-trivial due to the mixing of momentum-dependent even- and odd-parity superconducting states and a momentum-dependent SOC. To identify the individual roles of the SOC and the parity of the superconducting state, we first consider a simplified case where we assume a spin-singlet gap function and ignore the momentum dependence of the Ising SOC and the in-plane magnetic field. In this approximation, the resulting gap equation allows us to decouple the transition temperature-dependent contribution from the gap matrix. The resulting equation is given by
\begin{align}
    \text{ln}\left(\frac{T}{T_{\mathrm{c}}}\right)=\frac{-h_{x}^{2}}{h_{x}^{2}+\delta_{}^{2}}\Bigl\{\mathop{\mathrm{Re}}\Bigl[\psi\bigl(\frac{1}{2}+i\frac{\sqrt{h_{x}^{2}+\delta_{}^{2}}}{2\pi k_{B}T_{}}\bigr) \Bigr] -\psi\bigl(\frac{1}{2}\bigr)\Bigr\}.
    \label{eqn_toy_model}
\end{align}
The behavior of the above simplified model has been considered in previous works on monolayer TMDs \cite{ilic2017enhancement,yi2022crossover} 
and as shown in Fig.~\ref{fig:toy} leads to an enhancement of the upper critical field and a SOC-induced low-temperature upturn in the critical magnetic field. For odd-parity states, the transition temperature remains independent of magnetic fields. It is believed that a similar SOC-induced enhancement of the upper critical field explains the large upper critical fields in Ising superconductors compared to the Pauli limit. For example, monolayer TaS$_2$ with stronger Ising SOC compared to monolayer \nb~exhibits a correspondingly larger critical field~\cite{de2018tuning}. Additionally, the critical field is suppressed by increasing the number of layers in \nb~\cite{de2018tuning}. This effect can be related to the weakening of Ising SOC due to cancellation of oppositely oriented internal magnetic fields from consecutive layers.

\begin{figure}[]
    \centering
    \includegraphics[width=0.5\textwidth]{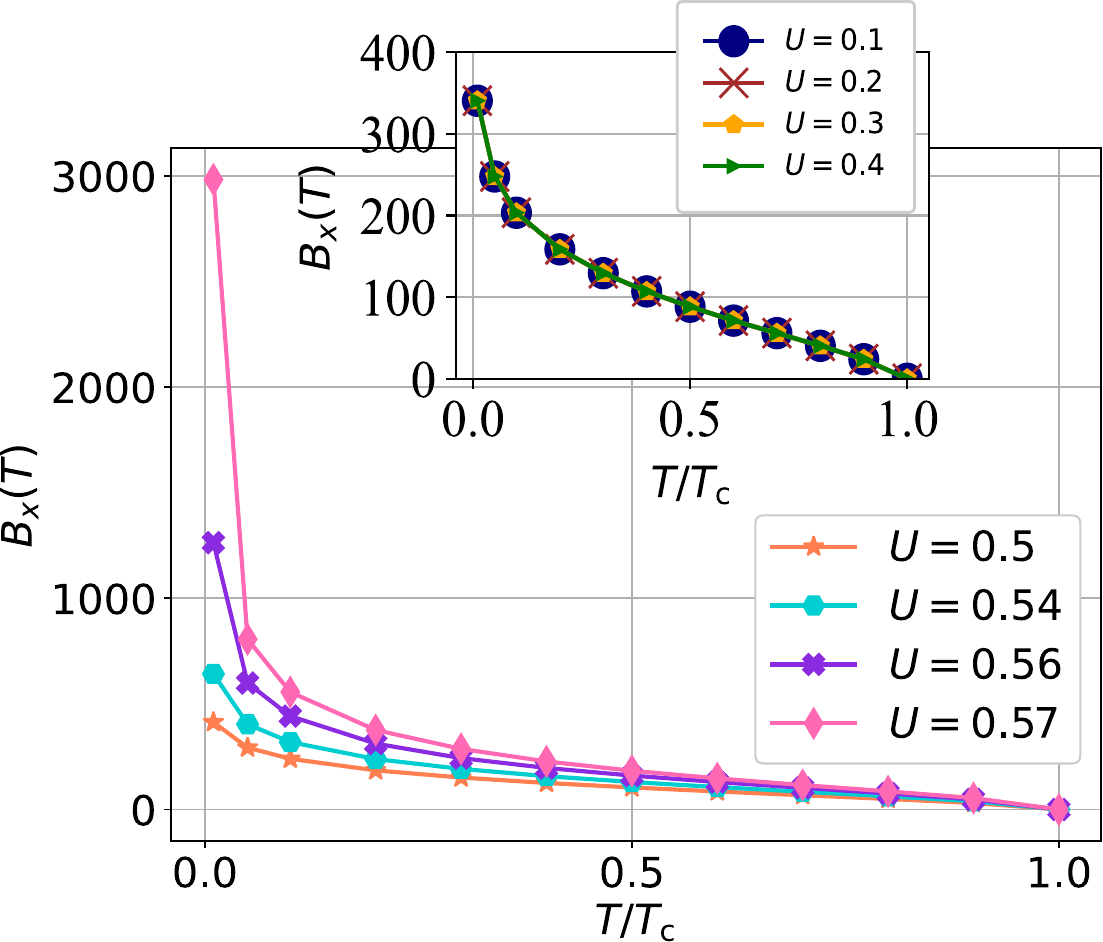}
    \caption{Effect of in-plane magnetic field on $T/T_{\mathrm{c}}$ in the presence of Ising SOC. Here, as opposed to the results shown in Fig.~\ref{fig:toy}, we have taken into account the full momentum-dependence of the Ising SOC and the allowed mixing of even- and odd-parity superconducting states.
    }
     \label{fig:in_plane_mag_with_ising}
\end{figure}

We next consider the numerical solution of the full linearized gap equation $(\ref{fullgapeqn})$ in the presence of momentum-dependent Ising SOC and mixed even- and odd-parity superconducting states. Note that with this generalization, the magnetic field and transition temperature cannot be decoupled from the gap matrix itself, and the self-consistent equation has to be numerically solved (see SM). Although the transition temperature is modulated by the magnetic field, the dominant gap function remains unchanged from the gap structure obtained in zero magnetic field.
As can be seen in Fig.~\ref{fig:in_plane_mag_with_ising}, the SOC significantly enhances the critical magnetic fields in the low-temperature limit with increasing $U$. For $U>0.6$\,eV the transition temperature becomes independent of the magnetic field, indicating a superconducting state that is dominated by the odd-parity channel. This result agrees with the gap function obtained for $U=1.0$\,eV shown in Fig.~\ref{fig_gap_ising}.

The low-temperature upturn in the second order transition line shown in Fig.~\ref{fig:in_plane_mag_with_ising} is much stronger than the upturn expected from a purely SOC effect considered in Fig.~\ref{fig:toy} for a spin-singlet system. The additional upturn is explained by the mixing of odd parity superconducting state. However, although a similar low-temperature upturn has been reported for Ising superconductors \cite{Zhang_2021} and explained to be a consequence of Ising SOC, a large enhancement in the critical field has not been reported for monolayer NbSe$_2$. The suppression of an upturn in the critical field of Ising superconductors can originate via inhomogeneities or through the effect of competing orders.
For example, it has been suggested that the presence of disorder can lead to an effective reduction in the SOC energy scale~\cite{liu2020microscopic}. Therefore, disorder effects would not only suppress the superconducting gap, but the reduced SOC can also reduce the low-temperature upturn in critical magnetic fields. Additionally, the low-temperature upturn could also be suppressed by the presence of an additional Rashba SOC \cite{liu2020microscopic,falson2020type} that could originate from a weak breaking of the in-plane mirror symmetry due to a substrate layer. A finite Rashba SOC contribution induces an in-plane spin component and reduces the protection of the superconducting state from in-plane magnetic fields. 

Another interesting feature relevant to monolayer \nb~is the underlying 3Q CDW order. Previous studies on this system have ignored the effect of CDW on the superconducting state since the CDW experimentally shows up only as a small anomaly in local DOS measurements, indicating the opening of an anisotropic CDW gap over small regions of the Fermi surface \cite{ugeda2016characterization,zhang2022visualization}. Additionally, suppression of CDW by disorder\cite{cossu2020strain,cossu2018unveiling} or pressure seems to have only a rather small effect on the superconducting transition temperature \cite{das2023electron,ugeda2016characterization}. First principles calculations indicate that the CDW gap primarily opens up at the Fermi surface around the $K$, $K'$ points \cite{Zheng2019}. However, if we note that these are also the regions where the momentum-dependent Ising SOC is large, the CDW will likely influence the superconducting pairing and lead to modification of low-temperature upturn in the critical magnetic field. Another important contribution of the underlying CDW phase could be to smear-out the two-gap feature obtained in the DOS.

Since the CDW state is described by a 3Q order, the system retains the point group symmetry but breaks the translational symmetry. One consequence of coupling of the superconducting and CDW order would be to make the SC order itself spatially inhomogenous and develop a periodicity similar to the CDW order. Although the nature of such a modulation would depend on the strength of the repulsive coupling between CDW and SC, similar effects have been seen in bulk 2H-NbSe2 \cite{liu2021discovery}. Additionally, the CDW can lead to unusual charge modulations near disorder sites\cite{debnath2024charge} such as the possibility of a stripe order near edge states\cite{zhang2022visualization}. It would be interesting to explore how the superconducting gap behaves near the edge in the presence of such stripe modulations of the CDW order parameter. For the homogeneous system, the CDW is partially gapping the Fermi surface, thus pairing strength is expected to decrease as seen in corresponding electron-phonon calculations\cite{zheng2019electron}. Similar effects are expected for spin-fluctuation mediated superconductivity; concrete pairing calculations with larger elementary cell in presence of CDW are beyond the scope of this work due to the large number of orbitals in such a calculation.

\section{\label{sec:level5}Summary}

In summary, we have provided a theoretical framework to compute the effective momentum-dependent superconducting pairing interaction from repulsive interactions in Ising superconductors. The pairing is obtained from charge- and spin-fluctuation diagrams and incorporates the full momentum-dependence of the band structure and the Ising SOC. While this formalism is not established as {\it the} method for pairing from repulsive interaction, it is nevertheless well studied and has successfully described $d_{x^2-y^2}$ superconductivity in cuprates and $s^\pm$ superconducting order in the iron-based superconductors~\cite{Scalapino2012,Mazin2008Unconventional,HIRSCHFELD2016197,Chubukovreview,Romer2020,Kreisel_review}.  We have applied this formalism to the case of monolayer \nb{} and explored the consequences of unconventional superconductivity in this material. We find that a reliable description of the preferred superconducting state depends crucially on the inclusion of all relevant Fermi pockets and the Ising SOC. Additionally, the solutions of the superconducting gap equation exhibit a significant dependence of the onsite Coulomb interaction, which determines the degree of mixing of even- and odd-parity superconducting states. While the odd-parity solutions exist as subdominant order parameters in the absence of Ising SOC, they become a dominant part of the leading gap structure in the presence of realistic amplitudes of Ising SOC. We have also studied the effect of the Hund’s coupling parameter $J$, revealing that an enhancement of $J$ results in a further dominance of the odd-parity channel in the mixing. These results highlight monolayer TMDs as interesting potential candidates for realising the interesting phenomenology of spin-triplet superconductivity, see also Refs.~\onlinecite{Ammar2023,Hsu2020}.

The fact that we find a leading two-dimensional Irrep is of direct relevant to several of the puzzling experiments mentioned in the Introduction. For example, the two-fold symmetric magnetoresistance~\cite{hamill2021two,Cho2022} could emerge if the imposed magnetic field breaks the degeneracy between the two leading solutions. Since these separately are two-fold symmetric, this is consistent with the experiments. Additionally, the unconventional nature of the pairing states make them fragile to disorder leading to a quick suppression of $T_c$. An initial increase of $T_c$, however, as seen in monolayer NbSe$_2$ with Si adatoms would require additional effects beyond the current modelling, possible related to wavefunction multifractality or suppression of competing CDW order~\cite{Zhao2019,Gastiasoro2018}.

The detailed momentum dependence of the gap structure obtained within the present approach, depends strongly on the inclusion of Ising SOC and the strength of the interactions. In the realistic cases, the $\Gamma$-centered pocket acquires significant gap and all pockets exhibit nodes. These nodes, however, are not obviously manifest in the resulting DOS since  the 2D Irreps will superpose and break time-reversal to gap out additional states. Additionally, STM measurements will preferentially tunnel into the $d_{z^2}$ orbital, highlighting only that part of the spectrum.    

Finally, we have studied the response of the preferred superconducting states to external magnetic fields. In the case of an out-of-plane magnetic field ignoring orbital effects, the transition temperature of the superconducting state features the expected behavior for a second-order phase transition. For in-plane magnetic fields, the SOC leads to a large upturn of the critical magnetic field at low temperatures which grows with increasing Coulomb interaction. For sufficiently large values of $U$, the solution is completely dominated by an odd-parity superconducting gap, and the transition temperature becomes independent of magnetic field strength. Our work lays the essential ground for a more comprehensive analysis of the superconducting state in Ising superconductors that can include the effect of electron-phonon interaction and CDW in addition to spin-fluctuation-mediated pairing contributions considered in this work. 
\begin{acknowledgments}
We acknowledge useful discussions with D. Agterberg, M. Khodas, I. Mazin, and R. Nanda. A.K.~acknowledges support by the Danish National Committee for Research Infrastructure (NUFI) through the ESS-Lighthouse Q-MAT. S. M., and S. R. acknowledge support from IIT Madras through HRHR travel mobility grant. 
\end{acknowledgments}

\bibliography{ref.bib}

\newpage

\textbf{SUPPLEMENTAL MATERIAL}
\appendix

\section{Band structure and Fermi surface}
\begin{figure}[h]
    \includegraphics[width=1.0\linewidth]{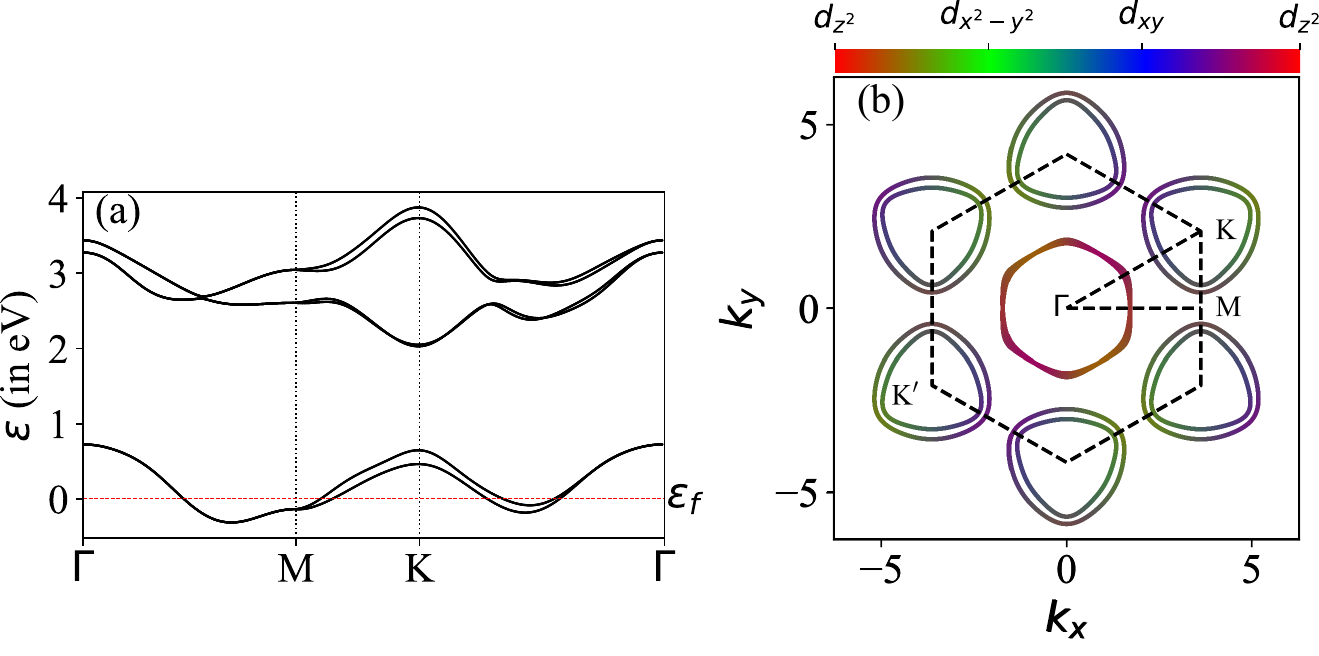}     
    \caption{(a) The band structure with full bandwith. (b) The Fermi surface colored according to the orbital content.}
    \label{fig_band_fs}
\end{figure}
Fig.~\ref{fig_band_fs}-(a) shows Ising SOC spilt bands. There are total 6-bands. We primarily focus on low energy bands crossing the Fermi surface. Fig.~\ref{fig_band_fs}-(b) shows orbital resolved Fermi surface. As can be seen in figure $\Gamma$-centered pocket is primarily dominated by $d_{z^{2}}$ orbital and the K-centered pockets are primarily dominated by degenerate $d_{x^{2}-y^{2}}$ and $d_{xy}$ orbitals.

\section{The effective pairing interaction}
Here we give a detailed derivation of the pairing kernel used in the main paper, Eqn.(4). The rotationally invariant interaction Hamiltonian is given by
\begin{eqnarray}
H_{\mathrm{int}}&=&\frac{U}{2} \sum_{i,\mu,\sigma} n_{i \mu \sigma}n_{i \mu \Bar{\sigma}}
+\frac{U'}{2} \sum_{i,\mu \neq \nu , \sigma} n_{i \mu \sigma }n_{i \nu \Bar{\sigma}}\nonumber\\
&+&\frac{U'-J}{2} \sum_{i,\mu \neq \nu , \sigma} n_{i \mu \sigma }n_{i \nu \sigma} \nonumber\\
&+&\frac{J}{2} \sum_{i,\mu \neq \nu}   \sum_{\sigma} \left[c^{+}_{i \mu \sigma} c^{+}_{i \nu \Bar{\sigma}} c^{}_{i \mu \Bar{\sigma}} c^{}_{i \nu \sigma}   \right]\nonumber\\
&+&\frac{J'}{2} \sum_{i,\mu \neq \nu}   \sum_{\sigma} \left[   c^{+}_{i \mu \sigma} c^{+}_{i \mu \Bar{\sigma}} c^{}_{i \nu \Bar{\sigma}} c^{}_{i \nu \sigma}   \right]
\end{eqnarray}
Here, the $U$ term, the $U'$ term, and the $J$ term denote the intra-orbital, inter-orbital Hubbard
repulsion and the Hund’s rule coupling as well as the pair hopping. 
We work in the spin-rotational invariant setting where the onsite Coulomb terms are related by $U = U' + 2J$ and $J=J'$.
Restricting to Cooper pair channel the above interaction Hamiltonian can be re-written with a compact notation  as following
\begin{align}
H_{\mathrm{int}}=\frac{1}{2} \sum_{\vk,\vk',\tilde{l}} [U]^{\tilde{l}_1,\tilde{l}_2}_{\tilde{l}_3,\tilde{l}_4} \, c^{\dagger}_{\vk,\tilde{l}_1} \,c^{\dagger}_{-\vk,\tilde{l}_3} \, c_{-\vk',\tilde{l}_2}\, c_{\vk',\tilde{l}_4}
\label{eqn:int_hamil}
\end{align}
\begin{figure}[tb]
   \begin{tikzpicture}[scale=0.7]
   \begin{feynman}
     \draw[black, very thick] (0,0) rectangle (5,2) node[pos=.5] {$[V(\textbf{k},\textbf{k}')]^{\tilde{l}_1,\tilde{\mu_2}}_{\tilde{l}_3,\tilde{l}_4}$};
\vertex (a) at (-1.5,4) {\(\textbf{k}\), \( \tilde{l}_1 \)};
\vertex (b) at (-1.5,-2) {\(\textbf{k}'\), \( \tilde{l}_4 \)};

\vertex (m) at (0.4,0) ;
\vertex (n) at (0.4,2) ;
\vertex (o) at (4.6,0) ;
\vertex (p) at (4.6,2) ;

\vertex (c) at (6.5,4) {-\(\textbf{k}\), \( \tilde{l}_3 \)};
\vertex (d) at (6.5,-2) {-\(\textbf{k}'\), \( \tilde{l}_2 \)};

\diagram* {
(b) -- [plain, with arrow=0.91, line width = 1.1pt] (m) ,
(n) -- [plain, with arrow=0.91, line width = 1.1pt] (a) ,
(d) -- [plain, with arrow=0.91, line width = 1.1pt] (o) ,
(p) -- [plain, with arrow=0.91, line width = 1.1pt] (c) 
};
\end{feynman}
   \end{tikzpicture}
    \caption{Effective pairinig vertex}
    \label{fig_pairing_vertex}
   \end{figure}
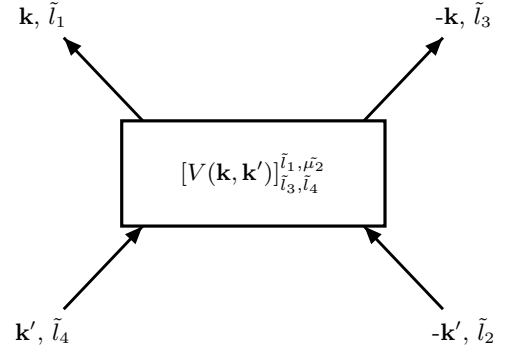
Here, $[\tilde{l}:=(l,\sigma)]$. The bare electron-electron interaction $ [U]^{\tilde{l}_1,\tilde{l}_2}_{\tilde{l}_3,\tilde{l}_4}$, i.e. the first order correction in $ [V]^{\tilde{l}_1,\tilde{l}_2}_{\tilde{l}_3,\tilde{l}_4}$ is given by\\

\begin{center}
\begin{tabular}{c c c }
$[U]_{\mu \Bar{s} \mu s}^{\mu s \mu \Bar{s}} = U$ & $[U]_{\mu \Bar{s} \nu s}^{\nu s \mu \Bar{s}}= U'$ & $[U]_{\mu \Bar{s} \nu s}^{\mu s \nu \Bar{s}}= J'$ \\ 
                                                &                                                                                                     \\
$[U]_{\nu \Bar{s} \nu s}^{\mu s \mu \Bar{s}} = J$ &\,\,\,\,\,\,\,\,\,\, $[U]_{\nu s \mu s}^{\mu s \nu s}= U'-J$           & \\
\\
 \,\,\,\,\,\,$[U]_{\mu \Bar{s} \mu \Bar{s}}^{\mu s \mu s} =-U $ &\,\,\,\,\,\,$[U]_{\mu \Bar{s} \mu \Bar{s}}^{\nu s \nu s}=-U'   $ &\,\,\, $[U]_{\mu \Bar{s} \nu \Bar{s}}^{\mu s \nu s}= -J'$\\ 
                                                                                                            \\
\,\,\,\,\,\,$[U]_{\nu \Bar{s} \mu \Bar{s}}^{\mu s \nu s} =-J $ &\,\,\,\,\,\,\,\,\,\,\,\,\,\,\,\,\,$[U]_{\nu s \nu s}^{\mu s \mu s}=-U'+J $ &  
\end{tabular}
\end{center}

For higher order correction in $ [V]^{\tilde{l}_1,\tilde{l}_2}_{\tilde{l}_3,\tilde{l}_4}$ we sum up all bubble and ladder diagrams to infinite order in $U$ (RPA).

\subsection{Bubble Diagrams}
\begin{figure}[h]

   \begin{tikzpicture}[scale=0.65]
    \begin{feynman}
\vertex (a) at (-2,2) {\(\textbf{k}\), \( \tilde{l}_1 \)};
\vertex (b) at (-2,-2) {\(\textbf{k}'\), \( \tilde{l}_4 \)};
\vertex (m) at (0,0) ;
\vertex (n) at (2,0) ;
\vertex (o) at (5,0) ;
\vertex (p) at (7,0) ;
\vertex (c) at (9,2) {-\(\textbf{k}\), \( \tilde{l}_3 \)};
\vertex (d) at (9,-2) {-\(\textbf{k}'\), \( \tilde{l}_2 \)};
\vertex (px) at (1.7,0.7) { \( \tilde{\nu}_2 \)};
\vertex (qx) at (1.7,-0.7) {\( \tilde{\nu}_1 \)};
\vertex (sx) at (5.3,0.7) { \( \tilde{\nu}_3 \)};
\vertex (tx) at (5.3,-0.7) {\( \tilde{\nu}_4 \)};
\diagram* {
(b) -- [plain, with arrow=0.91, line width = 1.1pt] (m) -- [plain, with arrow=0.99, line width = 1.3pt] (a),
(m) -- [photon] (n),
(n) -- [plain, with arrow=0.97, line width = 1.1pt, half right , edge label'=\(\textbf{p}\)] (o),
(o) -- [plain, with arrow=0.97, line width = 1.1pt, half right , edge label'=\(\textbf{p} + (\textbf{k}-\textbf{k}')\)] (n),
(o) -- [photon] (p),
(d) -- [plain, with arrow=0.91, line width = 1.1pt] (p) -- [plain, with arrow=0.99, line width = 1.1pt] (c)
};
\end{feynman}
   \end{tikzpicture}
    \caption{Second order Bubble diagram}
    \label{fig_bubble_diagram_2nd_order}
   \end{figure}
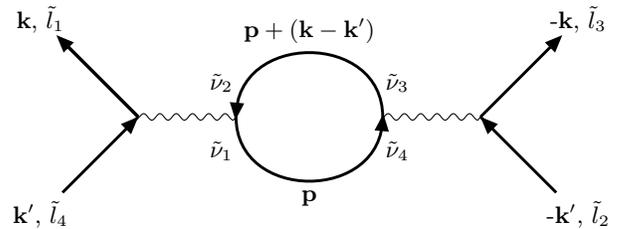
The second order bubble diagram shown in Fig.~(\ref{fig_bubble_diagram_2nd_order}) can be written as

\begin{align}
    U^{\tilde{l}_1,\tilde{\nu}_2}_{\tilde{\nu}_1,\tilde{l}_4}\, \,\chi^{\tilde{\nu}_1,\tilde{\nu}_2}_{\tilde{\nu}_3,\tilde{\nu}_4}\,\, U^{\tilde{\nu}_3,\tilde{l}_2}_{\tilde{l}_3,\tilde{\nu}_4}
\end{align}
with anti-commutation properties of operators, 
\begin{eqnarray}
   ( -U^{\tilde{l}_1,\tilde{l}_4}_{\tilde{\nu}_1,\tilde{\nu}_2})\, \,\chi^{\tilde{\nu}_1,\tilde{\nu}_2}_{\tilde{\nu}_3,\tilde{\nu}_4}\,\, (-U^{\tilde{\nu}_3,\tilde{\nu}_4}_{\tilde{l}_3,\tilde{l}_2})
\end{eqnarray}
In language of matrices this is simply
\begin{align}
    [U\,\chi\, U]^{\tilde{l}_1,\tilde{l}_4}_{\tilde{l}_3,\tilde{l}_2}
\end{align}
\begin{figure}[h]
    \begin{tikzpicture}[scale=0.55]
    \begin{feynman}

\vertex (a) at (-1,2) {\(\textbf{k}\), \( \tilde{l}_1\)};
\vertex (b) at (-1,-2){\(\textbf{k}'\), \( \tilde{l}_4 \)};
\vertex (m) at (0,0) {};
\vertex (n) at (2,0) {};
\vertex (o) at (5,0) {};
\vertex (p) at (7,0) {};
\vertex (p1) at (10,0) {};
\vertex (q1) at (12,0) {};
\vertex (c) at (13,2){-\(\textbf{k}\), \( \tilde{l}_3 \)};
\vertex (d) at (13,-2){-\(\textbf{k}'\), \( \tilde{l}_2\)};

\vertex (px) at (1.5,0.7) { \( \tilde{\nu}_2 \)};
\vertex (qx) at (1.7,-0.7) {\( \tilde{\nu}_1 \)};
\vertex (sx) at (5.3,0.7) { \( \tilde{\nu}_3 \)};
\vertex (tx) at (5.5,-0.7) {\( \tilde{\nu}_4 \)};

\vertex (ppx) at (6.5,0.7) { \( \tilde{\nu}_2' \)};
\vertex (qqx) at (6.7,-0.7) {\( \tilde{\nu}_1' \)};
\vertex (ssx) at (10.3,0.7) { \( \tilde{\nu}_3' \)};
\vertex (ttx) at (10.5,-0.7) {\( \tilde{\nu}_4' \)};

\diagram* {
(b) -- [plain, with arrow=0.97, line width = 1.1pt] (m) -- [plain, with arrow=0.97, line width = 1.1pt] (a),
(m) -- [photon] (n),
(n) -- [plain, with arrow=0.97, line width = 1.1pt ,half right , edge label'=\(\textbf{p}\) ] (o),
(o) -- [plain, with arrow=0.97, line width = 1.1pt ,half right , edge label'=\(\textbf{p} + (\textbf{k}-\textbf{k}')\)] (n),
(p) -- [plain, with arrow=0.97, line width = 1.1pt ,half right , edge label'=\(\textbf{p}'\) ] (p1),
(p1) -- [plain, with arrow=0.97, line width = 1.1pt ,half right, edge label'=\(\textbf{p}' + (\textbf{k}-\textbf{k}')\) ] (p),
(o) -- [photon] (p),
(p1) -- [photon] (q1),
(d) -- [plain, with arrow=0.97, line width = 1.1pt] (q1) -- [plain, with arrow=0.97, line width = 1.1pt] (c)
};

\end{feynman}
   \end{tikzpicture}
   \caption{Third order Bubble diagram}
   \label{fig_bubble_diagram_3rd_order}
\end{figure}
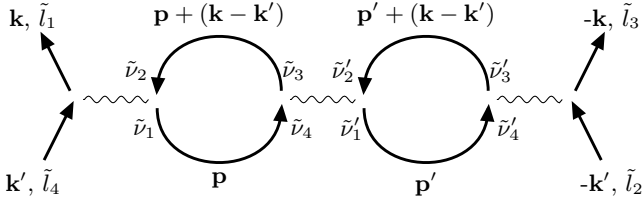
The third order bubble diagram shown in Fig.~(\ref{fig_bubble_diagram_3rd_order}) is given as
\begin{align}
   ( -U^{\tilde{l}_1,\tilde{l}_4}_{\tilde{\nu}_1,\tilde{\nu}_2})\, \,\chi^{\tilde{\nu}_1,\tilde{\nu}_2}_{\tilde{\nu}_3,\tilde{\nu}_4}\,\, 
    (- U^{\tilde{\nu}_3,\tilde{\nu}_4}_{\tilde{\nu}_1',\tilde{\nu}_2'})\, \,
     \chi^{\tilde{\nu}_1',\tilde{\nu}_2'}_{\tilde{\nu}_3',\tilde{\nu}_4'}\,\, 
   (- U^{\tilde{\nu}_3',\tilde{\nu}_4'}_{\tilde{l}_3,\tilde{l}_2}),
\end{align}
or un matrix notation
\begin{align}
   - [U\,\chi\, U\,\chi\,U]^{\tilde{l}_1,\tilde{l}_4}_{\tilde{l}_3,\tilde{l}_2}.
\end{align}
So, if we consider all orders of bubble diagrams, we get terms alternating in sign, a $n_{\text{th}}$ order term having the sign $(-1)^n$.\\
Thus the total contribution from all the bubble diagrams can be represented as following summation of all the orders in bubble
\newpage
\begin{widetext} 
\begin{eqnarray}
    -[V^\mathrm{{Total}}_{\mathrm{bub}}]^{\tilde{l}_1,\tilde{l}_2}_{\tilde{l}_3,\tilde{l}_4}&=&\left[(-U)\,\chi\,(-U)\right]^{\tilde{l}_1,\tilde{l}_4}_{\tilde{l}_3,\tilde{l}_2} -
    \left[(-U)\,\chi\,(-U)\,\chi\,(-U)\right]^{\tilde{l}_1,\tilde{l}_4}_{\tilde{l}_3,\tilde{l}_2}\nonumber+
      \left[(-U)\,\chi\,(-U)\,\chi\,(-U)\,\chi\,(-U)\right]^{\tilde{l}_1,\tilde{l}_4}_{\tilde{l}_3,\tilde{l}_2} -\ldots\nonumber\\
      &=&\left[\frac{(-U)\,\chi\,(-U)}{1+\chi\,(-U)}\right]^{\tilde{l}_1,\tilde{l}_4}_{\tilde{l}_3,\tilde{l}_2}
\end{eqnarray}
\end{widetext}

\subsection{Ladder Diagrams}
Lets, bring our attention to the ladders now
\begin{figure}[h]
  
   \begin{tikzpicture}[scale=0.6]
    \begin{feynman}

\vertex (a) at (0,0)  { \( \tilde{l}_2 \)};
\vertex (b) at (4,0){};
\vertex (c) at (10,0) {};
\vertex (d) at (14,0)  { \( \tilde{l}_3 \)};
\vertex (e) at (4,3) {};
\vertex (f) at (10,3) {};
\vertex (g) at (4,6) {\(\textbf{k}'\), \( \tilde{l}_4 \)};
\vertex (h) at (10,6) { \(\textbf{k}\), \( \tilde{l}_1 \)};

\vertex (px) at (4.4,0.3) { \( \tilde{\nu}_1 \)};
\vertex (qx) at (4.4,2.7) { \( \tilde{\nu}_2 \)};
\vertex (sx) at (9.6,0.3) { \( \tilde{\nu}_4 \)};
\vertex (tx) at (9.6,2.7) { \( \tilde{\nu}_3 \)};

\diagram* {
(a) -- [plain, with arrow=0.99, line width = 1.3pt, edge label=\(\textbf{-k}'\) ] (b) -- [ fermion, line width = 1.3pt, edge label=\(\textbf{p}\)] (c) -- [ plain, with arrow=0.99, line width = 1.3pt, edge label=\(\textbf{-k}\)] (d),
(f) -- [fermion , , line width = 1.3pt, edge label=\(\textbf{p} + (\textbf{k}+\textbf{k}')\)] (e),
(e) -- [plain, with arrow=0.99, line width = 1.3pt] (h),
(g) -- [plain, with arrow=0.99, line width = 1.3pt] (f),
(e) -- [photon] (b),
(f) -- [photon] (c)

};

\end{feynman}
   \end{tikzpicture}
 \caption{Second order Ladder diagram}
 \label{fig_ladder_diagram_2nd_order}
\end{figure}
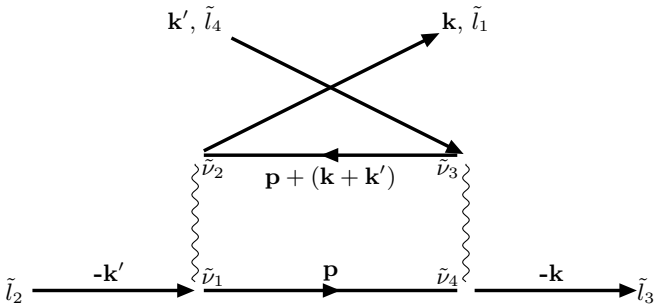
Fig.~(\ref{fig_ladder_diagram_2nd_order}) with anti-commutation property of operators can be written as
\begin{eqnarray}
 U^{\tilde{l}_1,\tilde{l}_2}_{\tilde{\nu}_1,\tilde{\nu}_2}\, \,\chi^{\tilde{\nu}_1,\tilde{\nu}_2}_{\tilde{\nu}_3,\tilde{\nu}_4}\,\, U^{\tilde{\nu}_3,\tilde{\nu}_4}_{\tilde{l}_3,\tilde{l}_4}=  [U\,\chi\, U]^{\tilde{l}_1,\tilde{l}_2}_{\tilde{l}_3,\tilde{l}_4}
\end{eqnarray}
\begin{figure}
    \begin{tikzpicture}[scale=0.5]
     \begin{feynman}

\vertex (a) at (0,0)  { \( \tilde{l}_2 \)};
\vertex (b) at (2,0){};
\vertex (bc) at (8,0){};
\vertex (c) at (14,0) {};
\vertex (d) at (16,0)  { \( \tilde{l}_3 \)};

\vertex (e) at (2,3) {};
\vertex (ef) at (8,3) {};
\vertex (f) at (14,3) {};

\vertex (g) at (2,6) { \(\textbf{k}'\), \( \tilde{l}_4 \)};
\vertex (h) at (14,6) { \(\textbf{k}\), \( \tilde{l}_1 \)};

\vertex (ex) at (7,0) {};
\vertex (fx) at (7,3) {};

\vertex (px) at (2.4,0.4) { \( \tilde{\nu}_1 \)};
\vertex (qx) at (2.4,2.6) { \( \tilde{\nu}_2 \)};
\vertex (sx) at (13.6,0.4) { \( \tilde{\nu}_4 \)};
\vertex (tx) at (13.6,2.6) { \( \tilde{\nu}_3 \)};

\vertex (ppx) at (8.5,0.4) { \( \tilde{\nu}_4' \)};
\vertex (qqx) at (8.5,2.6) { \( \tilde{\nu}_3' \)};
\vertex (ssx) at (7.5,0.4) { \( \tilde{\nu}_1' \)};
\vertex (ttx) at (7.5,2.6) { \( \tilde{\nu}_2' \)};

\diagram* {
(a) -- [plain, with arrow=0.99, line width = 1.3pt, edge label=\(\textbf{-k}'\)] (b) -- [ fermion, line width = 1.3pt, edge label=\textbf{p} ] (bc) -- [ fermion, line width = 1.3pt,  edge label=\(\textbf{p}'\) ] (c) -- [ plain, with arrow=0.99, line width = 1.3pt, edge label=\(\textbf{-k}\)] (d),
(f) -- [fermion, line width = 1.3pt, edge label=\(\textbf{p}' + (\textbf{k}+\textbf{k}')\)] (ef),
(ef) -- [fermion , line width = 1.3pt, edge label=\(\textbf{p} + (\textbf{k}+\textbf{k}')\)] (e),
(e) -- [plain, with arrow=0.99 , line width = 1.3pt] (h),
(g) -- [plain, with arrow=0.99, line width = 1.3pt] (f),
(e) -- [photon] (b),
(f) -- [photon] (c),
(ef) -- [photon] (bc)

};
\end{feynman}
   \end{tikzpicture}
   \caption{Third order Ladder diagram}
  \label{fig_ladder_diagram_3rd_order}
\end{figure}
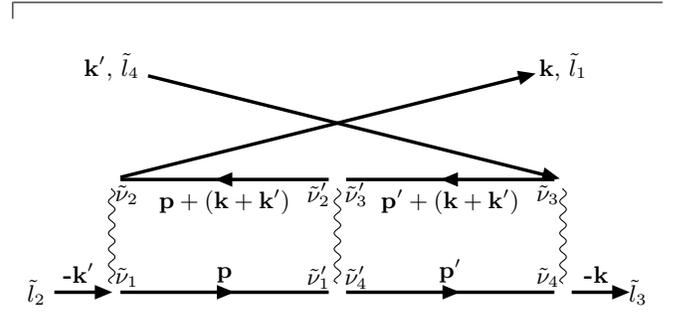
For, the third order ladder diagram in Fig.~(\ref{fig_ladder_diagram_3rd_order}) we have
\begin{align}
    U^{\tilde{l}_1,\tilde{l}_2}_{\tilde{\nu_1},\tilde{\nu}_2}\, 
    \,\chi^{\tilde{\nu}_1,\tilde{\nu}_2}_{\tilde{\nu}_2',\tilde{\nu}_1'}\,\, 
     U^{\tilde{\nu}_2',\tilde{\nu}_1'}_{\tilde{\nu}_4',\tilde{\nu}_3'}\, \,
     \chi^{\tilde{\nu}_4',\tilde{\nu}_3'}_{\tilde{\nu}_3,\tilde{\nu}_4}\,\, 
    U^{\tilde{\nu}_3',\tilde{\nu}_4}_{\tilde{l}_3,\tilde{l}_4}
    = [U\,\chi\, U\,\chi\,U]^{\tilde{l}_1,\tilde{l}_2}_{\tilde{l}_3,\tilde{l}_4}
\end{align}
Like bubble diagrams we will be doing summation in all orders of ladder diagrams. Few things to note here which is different in summation of ladder diagrams than summation in bubble diagrams As there is no fermionic looping, the susceptibility expression appears with a ``$-$" sign in front.There is no alternating signs appearing in series of ladder diagrams so, the denominator appears with a "$-$" sign.
\begin{widetext}
\begin{eqnarray}
    -[V^\mathrm{{Total}}_{\mathrm{ladder}}]^{\tilde{l}_1,\tilde{l}_2}_{\tilde{l}_3,\tilde{l}_4}&=&\left[(-U)\,(-\chi)\,(-U)\right]^{\tilde{l}_1,\tilde{l}_4}_{\tilde{l}_3,\tilde{l}_2} + \left[(-U)\,(-\chi)\,(-U)\,(-\chi)\,(-U)\right]^{\tilde{l}_1,\tilde{l}_4}_{\tilde{l}_3,\tilde{l}_2} \nonumber\\
    &+&\left[(-U)\,(-\chi)\,(-U)\,(-\chi)\,(-U)\,(-\chi)\,(-U)\right]^{\tilde{l}_1,\tilde{l}_4}_{\tilde{l}_3,\tilde{l}_2} \ldots\nonumber\\
      &=&\left[\frac{(-U)\,(-\chi)\,(-U)}{1-(-\chi)\,(-U)}\right]^{\tilde{l}_1,\tilde{l}_4}_{\tilde{l}_3,\tilde{l}_2}
\end{eqnarray}
From the preceding discussion, in presence of spin orbit coupling the final interaction vertex can be written as

\begin{eqnarray}
  [V(\vk,\vk')]^{\tilde{l}_1,\tilde{l}_2}_{\tilde{l}_3,\tilde{l}_4}&=&[U]^{\tilde{l}_1,\tilde{l}_2}_{\tilde{l}_3,\tilde{l}_4}-[\frac{U\,\chi_{0}\,U}{1-\chi_{0}\,U}]^{\tilde{l}_1,\tilde{l}_4}_{\tilde{l}_3,\tilde{l}_2}(\vk-\vk')+[\frac{U\,\chi_{0}\,U}{1-\chi_{0}\,U}]^{\tilde{l}_1,\tilde{l}_2}_{\tilde{l}_3,\tilde{l}_4}(\vk+\vk')
  \label{eqn:final_vertex}
\end{eqnarray}
This is the pairing vertex used in Eqn.(3) of the main text in the superconducting Hamiltonian,
\begin{eqnarray}
H_{\rm int}=\frac{1}{2} \sum_{\vk,\vk',\tilde{l}} [V]^{\tilde{l}_1,\tilde{l}_2}_{\tilde{l}_3,\tilde{l}_4}(\vk,\vk') \, c^{\dagger}_{\vk,\tilde{l}_1} \,c^{\dagger}_{-\vk,\tilde{l}_3} \, c_{-\vk',\tilde{l}_2}\, c_{\vk',\tilde{l}_4}.
\label{eqn:int_hamil}
\end{eqnarray}

\section{Gap equation for multiband superconductors}
\subsection{In absence of magnetic field}
In the following we derive the Gorkov equations and superconducting gap equation for a generic multi band Tight Binding Hamiltonian. The superconducting Hamiltonian is converted from the orbital basis to the band basis using the unitary transformation given in Eqn.(2) of the main text. For ease of notation we have changed the convention of writing down the pair potential in orbital basis, $ [V]^{\tilde{l}_1,\tilde{l}_2}_{\tilde{l}_3,\tilde{l}_4}$ in equation \ref{eqn:hsc} is same as $ [V]^{\tilde{l}_3,\tilde{l}_1}_{\tilde{l}_4,\tilde{l}_2}$ in equation \ref{eqn:final_vertex}. Note that in the following the combined notation $\tilde{l}_i=(l_i,\sigma_i)$. Assuming a homogeneous superconducting state, the Hamiltonian for a multi band superconductor can be expressed as,

\begin{eqnarray}
H&=&H_0+H_{\mathrm{sc}}\\
H_0&=&\sum_{\vk n \sigma}\epsilon_{n\sigma}(\vk)a^{\dag}_{\vk n \sigma}a_{\vk n \sigma}\label{eqn:h0}\\
H_{\mathrm{sc}}&=&\frac{1}{2}\sum_{n_i\sigma_i \vk \vk'} V^{n_1,n_2,\sigma_1,\sigma_2}_{n_3,n_4,\sigma_3,\sigma_4}(\vk,\vk')a^{\dag}_{-\vk n_1 \sigma_1}a^{\dag}_{\vk n_2 \sigma_2}a_{\vk' n_3 \sigma_3}a_{-\vk' n_4 \sigma_4}\label{eqn:hsc}
\end{eqnarray}
Here,
\begin{eqnarray}
   V^{n_1,n_2,\sigma_1,\sigma_2}_{n_3,n_4,\sigma_3,\sigma_4}(\vk,\vk')=\sum_{\{l_1,l_2,l_3,l_4\}} u^{n_{1}*}_{l_{1},\sigma_{1}}(-\vk)u^{n_{3}*}_{l_{3},\sigma_{3}}(\vk)V^{l_1,l_2,\sigma_1,\sigma_2}_{l_3,l_4,\sigma_3,\sigma_4}(\vk,\vk')u^{n_{2}}_{l_{2},\sigma_{2}}(\vk')u^{n_{4}}_{l_{4},\sigma_{4}}(-\vk')
\end{eqnarray}
where the pairing vertex on the left hand side of the above equation is in the band basis whereas it is expressed in the orbital basis on the right hand side. The above Hamiltonian is written in a multi band basis where $a_{\vk n \sigma}$($a_{\vk n \sigma}^\dagger$) annihilates (creates) an electron in band $n$, at momenta $\vk$, and quantum number $\sigma$. If the non interacting Hamiltonian preserves the $S_z$ quantum number then we can identify $\sigma$ with the true spin, $\sigma=(\uparrow,\downarrow)$. This would be true for example either in the absence of a spin orbit coupling or presence of a spin orbit coupling that does not generate off diagonal matrix elements (spin flip contributions). In the presence of spin flip contributions $\sigma$ would represent a pseudo spin index.

In the following calculation we will be studying the above Hamiltonian without assuming any specific form of non interacting Hamiltonian.\\
Let us define the generic Green's functions relevant to the problem as,
\begin{eqnarray}
 G^{\sigma,\sigma'}_{n,n'}(\vk,\tau_{1};\vk',\tau_{2})&=&- \bigl\langle T_{\tau} a_{\vk,n,\sigma}(\tau_{1})a^{\dagger}_{\vk',n',\sigma'}(\tau_{2}) \bigr\rangle \label{eqn:g}\\
   F^{\sigma,\sigma'}_{n,n'}(\vk,\tau_{1};\vk',\tau_{2})&=& \bigl\langle T_{\tau} a_{\vk,n,\sigma}(\tau_{1})a^{}_{-\vk',n',\sigma'}(\tau_{2}) \bigr\rangle \label{eqn:f}\\
   F^{* \sigma,\sigma'}_{n,n'}(\vk,\tau_{1};\vk',\tau_{2})&=& \bigl\langle T_{\tau} a^{\dagger}_{-\vk,n,\sigma}(\tau_{1})a^{\dagger}_{\vk',n',\sigma'}(\tau_{2}) \label{eqn:f*}\bigr\rangle
\end{eqnarray}

Here $T_{\tau}$ is the time time ordering operator in the imaginary time ($\tau$) formalism. After a mean field decomposition, the superconducting gap equation can be defined as,
\begin{eqnarray}
 \Delta^{n_{2},n_{1}}_{\sigma_{2},\sigma_{1}}(\vk)&=&-\sum_{\vk',n_{3},n_{4},\sigma_{3},\sigma_{4}}V^{n_1,n_2,\sigma_1,\sigma_2}_{n_3,n_4,\sigma_3,\sigma_4}(\vk,\vk')\bigl\langle a_{\vk',n_{3},\sigma_{3}}(\tau)a^{}_{-\vk',n_{4},\sigma_{4}}(\tau) \bigr\rangle  \\
 \Delta^{\ast n_{2},n_{1}}_{\sigma_{2},\sigma_{1}}(\vk)&=&-\sum_{\vk',n_{3},n_{4},\sigma_{3},\sigma_{4}}V^{n_1,n_2,\sigma_1,\sigma_2}_{n_3,n_4,\sigma_{3},\sigma_{4}}(\vk',\vk)\bigl\langle  a^{\dagger}_{-\vk',n_{3},\sigma_{3}}(\tau)a^{\dagger}_{\vk',n_{4},\sigma_{4}}(\tau)\bigr\rangle 
\end{eqnarray}
In the above the we do not assume a particular mechanism for the origin of superconducting pairing potential. The pair potential can be considered to be given by matrix elements given by,
\begin{equation}
\label{eqn:v}
 V^{n_1,n_2,\sigma_1,\sigma_2}_{n_3,n_4,\sigma_3,\sigma_4}(\vk,\vk') = \bigl\langle -\vk,n_{1},\sigma_{1};\vk,n_{2},\sigma_{2} \bigl|\hat{V}\bigr| -\vk',n_{4},\sigma_{4}; \vk',n_{3},\sigma_{3}\bigr\rangle
\end{equation}

Equation \ref{eqn:v} has the following symmetry properties

\begin{eqnarray}
\left.
\begin{aligned}
\label{eqn:vproperty}
V^{n_1,n_2,\sigma_1,\sigma_2}_{n_3,n_4,\sigma_3,\sigma_4}(\vk,\vk')
=V^{n_2,n_1,\sigma_2,\sigma_1}_{n_4,n_3,\sigma_4,\sigma_3}(-\vk,-\vk')
=V^{n_2,n_1,\sigma_2,\sigma_1}_{n_3,n_4,\sigma_3,\sigma_4}(-\vk,\vk')
=V^{n_1,n_2,\sigma_1,\sigma_2}_{n_4,n_3,\sigma_4,\sigma_3}(\vk,-\vk')
=V^{n_4,n_3,\sigma_4,\sigma_3}_{n_2,n_1,\sigma_2,\sigma_1}(\vk',\vk)
\end{aligned}
\right\}
\end{eqnarray}

In the following we use the equation of motion for the Greens functions to derive the Gorkov equation's, and the following anti-commutation property of the operators, \begin{eqnarray}
 \left[ \hat{A} \hat{B} , \hat{C} \right] &=& \hat{A} \{\hat{B},\hat{C}\} -\{\hat{A},\hat{C}\} \hat{B} \label{eqn:commutation1} \\
 \left[ \hat{A} \hat{B} \hat{C} \hat{D} , \hat{E}\right] &=&  \hat{A} \hat{B}[\hat{C} \hat{D} ,\hat{E}]+
 [\hat{A} \hat{B}, \hat{E} ]  \hat{C} \hat{D}\\
 \left[ \hat{A} \hat{B} \hat{C} \hat{D} , \hat{E}\right] &=& \hat{A} \hat{B} \hat{C} \{\hat{D},\hat{E}\} - \hat{A} \hat{B} \{\hat{C},\hat{E}\} \hat{D} + \hat{A} \{\hat{B},\hat{E}\}
\hat{C} \hat{D} - \{\hat{A},\hat{E}\}  \hat{B} \hat{C} \hat{D} \label{eqn:commutation2}
\end{eqnarray}

For the particle hole Greens' function we get (using Eq.~(\ref{eqn:g}))

\begin{eqnarray}
    \frac{\partial G^{\sigma,\sigma'}_{n,n'}}{\partial \tau}
    = - \frac{\partial \bigl\langle T_{\tau} a_{\vk,n,\sigma}(\tau)a^{\dagger}_{\vk',n',\sigma'}(0) \bigr\rangle}{\partial \tau}
     =-\delta(\tau) \delta_{n,n'} \delta_{\vk,\vk'} \delta_{\sigma,\sigma'}-\bigl\langle T_{\tau} \left[H,a_{\vk,n,\sigma}(\tau)\right]a^{\dagger}_{\vk',n',\sigma'}(0) \bigr\rangle \label{eqn:hg}
\end{eqnarray}

The commutator with the Hamiltonian involves $\left[H_{0}(\tau)+H_{\mathrm{sc}}(\tau),a_{\vk,n,\sigma}(\tau)\right]$. Using Eq.~(\ref{eqn:h0}) we obtain
\begin{eqnarray}
    \left[H_{0}(\tau),a_{\vk,n,\sigma}(\tau)\right]&=& - \epsilon_{n \sigma} (\vk) a_{\vk n \sigma}(\tau)\label{eqn:h0g}.
\end{eqnarray}
In the above we utilize the anti commutation relationships between the Fermion operators. Similarly, from Eq~\ref{eqn:hsc}

\begin{eqnarray}
  \left[H_{\mathrm{sc}},a_{\vk,\tilde{m}}(\tau)\right]
     &=& \sum_{\tilde{m}_i, \vs} V^{\tilde{m}_1,\tilde{m}}_{\tilde{m}_3,\tilde{m}_4}(\vk ,\vs )
     a^{\dag}_{-\vk \tilde{m}_1}(\tau)a_{\vs \tilde{m}_3}(\tau)a_{-\vs \tilde{m}_4}(\tau) \label{eqn:hscg}     
\end{eqnarray}
Note that in the above equation we have introduced a compact notation $\tilde{m}_i=(n_i,\sigma_i)$. Replacing the results from Eq~\ref{eqn:h0g}, and Eq~\ref{eqn:hscg} into Eq~\ref{eqn:hg} we get

\begin{eqnarray}
    \frac{\partial G_{\tilde{m},\tilde{m}'}}{\partial \tau}
    &=&-\delta(\tau)\delta_{\tilde{m},\tilde{m}'} \delta_{\vk,\vk'}+
    \epsilon_{\tilde{m}} (\vk) \bigl\langle T_{\tau} a_{\vk \tilde{m}}(\tau)a^{\dagger}_{\vk' \tilde{m}'}(0) \bigr\rangle \nonumber \\
    &&-\sum_{\substack{\tilde{m}_1 \rightarrow \tilde{m}_3 \\\vp }}  V^{\tilde{m}_1,\tilde{m}}_{\tilde{m}_2,\tilde{m}_3}(\vk ,\vp ) \bigl\langle T_{\tau} a^{\dag}_{-\vk \tilde{m}_1}(\tau)a_{\vp \tilde{m}_2}(\tau)a_{-\vp \tilde{m}_3}(\tau)a^{\dagger}_{\vk' \tilde{m}' }(0) \bigr\rangle
\end{eqnarray}   

Now performing mean field decomposition

\begin{eqnarray}    
 \frac{\partial G_{\tilde{m},\tilde{m}'}}{\partial \tau}(\vk,\tau;\vk',0)
    =-\delta(\tau)\delta_{\tilde{m},\tilde{m}'} \delta_{\vk,\vk'} -\epsilon_{\tilde{m}} (\vk) G_{\tilde{m},\tilde{m}'}(\vk,\tau;\vk',0)
    +\sum_{\tilde{m}_1}  \Delta^{\tilde{m},\tilde{m}_{1}}(\vk) F^{\ast}_{\tilde{m}_1,\tilde{m}'}(\vk,\tau;\vk',0) \label{eqn:gform}
\end{eqnarray}

Here, $\Delta^{\tilde{m}_{},\tilde{m}_{1}}(\vk)=-\sum_{\vp,\tilde{m}_{2},\tilde{m}_{3}}V^{\tilde{m}_{1},\tilde{m}_{}}_{\tilde{m}_2,\tilde{m}_3}(\vk,\vp)\bigl\langle a_{\vp,\tilde{m}_2}(\tau)a_{-\vp,\tilde{m}_{3}(\tau)} 
\bigr\rangle $. We can do the same procedure for $F$ and $F^{\ast}$ and 
use the transformation to Matsubara frequencies $G_{\tilde{m},\tilde{m}'}(\vk,\tau;\vk',0)= k_{\mathrm{B}}T \sum_{n}   G_{\tilde{m},\tilde{m}'}(\vk,\vk'; \omega_{n}) \exp{(-i \, \omega_{n} \tau)}$, this leads to following Gorkov equations
\begin{eqnarray}
(i \omega_{n} - \epsilon_{\tilde{m}}(\vk)) G^{\tilde{m},\tilde{m}'}(\vk,\vk'; \omega_{n})+
\sum_{\tilde{m}_1 }  \Delta^{\tilde{m},\tilde{m}_{1}}(\vk) F^{\ast \tilde{m}_{1},\tilde{m}'}(\vk,\vk'; \omega_{n})&=&\delta_{\vk,\vk'} \delta_{\tilde{m},\tilde{m}'}\\
(i \omega_{n} - \epsilon_{\tilde{m},\sigma}(\vk)) F^{\tilde{m},\tilde{m}'}(\vk,\vk'; \omega_{n})-
\sum_{\tilde{m}_1 }  \Delta^{\tilde{m},\tilde{m}_1}(\vk) G^{\tilde{m}',\tilde{m}_1}(-\vk',-\vk; -\omega_{n})&=&0\\
(i\omega_{n} + \epsilon_{m}(-\vk)) F^{\ast m,m'}(\vk,\vk'; \omega_{n})+
\sum_{m_1  }  \Delta^{\ast m_1,m}(\vk) G^{m_{1},m'}(\vk,\vk'; \omega_{n})&=&0
\end{eqnarray}

Then for translationally invariant system $G^{m,m'}_{}(\vk,\vk'; \omega_{n}) =  G^{m,m'}_{}(\vk,\omega_{n}) \delta(\vk-\vk')$.

The Gorkov equations reduce to
\begin{eqnarray}
(i \omega_{n} - \epsilon_{\tilde{m}}(\vk)) G^{\tilde{m},\tilde{m}'}(\vk,\omega_{n}) +
\sum_{m_1 }  \Delta^{\tilde{m},\tilde{m}_{1}}(\vk) F^{\ast \tilde{m}_{1},\tilde{m}'}(\vk,\omega_{n}) &=&\delta_{m,m'} \label{eqn:gorkov1}\\
(i\omega_{n} - \epsilon_{m}(\vk)) F^{\tilde{m},\tilde{m}'}(\vk,\omega_{n}) -
\sum_{m_1  }  \Delta^{\tilde{m},\tilde{m}_{1}}(\vk) G^{\tilde{m}',\tilde{m}_1}(-\vk,-\omega_{n}) &=&0\label{eqn:gorkov2}\\
(i \omega_{n} + \epsilon_{m}(-\vk)) F^{\ast \tilde{m},\tilde{m}'}_{}(\vk,\omega_{n}) +
\sum_{\tilde{m}_1 }  \Delta^{\ast \tilde{m}_1,\tilde{m}}(\vk) G^{\tilde{m}_{1},\tilde{m}'}(\vk,\omega_{n}) &=&\label{eqn:gorkov3}0,
\end{eqnarray}
where
\begin{eqnarray}
\label{eqn:deltaform}
 \Delta^{\tilde{m}_{2},\tilde{m}_{1}}_{}(\vk)&=&-\sum_{\substack{n, \tilde{m}_3 , \tilde{m}_4 ,  \\\vk }}  V^{\tilde{m}_1,\tilde{m}_2}_{\tilde{m}_3,\tilde{m}_4}(\vk,\vk')F^{\tilde{m}_{3},\tilde{m}_{4}}_{}(\vk',\omega_{n}).
\end{eqnarray}
To evaluate the linearized gap equation we consider F to be linear order in $\Delta$. This is achieved by replacing $G \rightarrow G^{0}$.\\
Then from Eq.~(\ref{eqn:gorkov2}) we get
\begin{eqnarray}
 F^{\tilde{m},\tilde{m}'}_{}(\vk,\omega_{n}) = \sum_{\tilde{m}_1} \frac{ \Delta^{\tilde{m},\tilde{m}_{1}}(\vk) G^{(0) \tilde{m}',\tilde{m}_1}(-\vk,-\omega_{n}) }{(i\omega_{n} - \epsilon_{\tilde{m}}(\vk)) }  
\end{eqnarray}
and from Eq.~(\ref{eqn:gorkov1})
\begin{eqnarray}
 G^{(0) \tilde{m}',\tilde{m}_1}(-\vk,-\omega_{n}) = -\frac{ \delta_{\tilde{m}',\tilde{m}_1}}{(i \omega_{n} + \epsilon_{\tilde{m}'}(-\vk)) }  
\end{eqnarray}

Combining the previous two equations we get
\begin{eqnarray}
 F^{\tilde{m},\tilde{m}'}(\vk,\omega_{n}) &=& -\sum_{\tilde{m}_1} \frac{ \Delta^{\tilde{m},\tilde{m}_{1}}(\vk) \delta_{\tilde{m}',\tilde{m}_1}  }{(i\omega_{n} - \epsilon_{\tilde{m}}(\vk))(i\omega_{n} + \epsilon_{\tilde{m}'}(-\vk)) } \nonumber \\
  &=& -\frac{ \Delta^{\tilde{m},\tilde{m}'}_{}(\vk)  }{(i\omega_{n} - \epsilon_{\tilde{m}}(\vk))(i\omega_{n} + \epsilon_{\tilde{m}'}(-\vk)) }  
\end{eqnarray}

Putting the previous equation back to \ref{eqn:deltaform} we get the linearized gap equation for a generic multi band homogeneous system
\begin{eqnarray}
\label{eqn:gapform}
 \Delta^{\tilde{m}_{1},\tilde{m}_{2}}_{}(\vk)&=&\sum_{\substack{n, \tilde{m}_3 , \tilde{m}_4 ,\\\vk' }}  \frac{V^{\tilde{m}_2,\tilde{m}_1}_{\tilde{m}_3,\tilde{m}_4}(\vk,\vk')\Delta^{\tilde{m}_{3},\tilde{m}_{4}}_{}(\vk')}{(i\omega_{n} - \epsilon_{\tilde{m}_3}(\vk'))(i\omega_{n} + \epsilon_{\tilde{m}_4}(-\vk')) } 
\end{eqnarray}
For system with Ising SOC Let us assume the $(\vk,\vk')$ represent the wave vectors for two bands that form the Fermi surface in the presence of Ising spin orbit coupling. Then we have in Eq~\ref{eqn:gapform}

\begin{eqnarray}
\left.
\begin{aligned}
\tilde{m}_1 \neq \tilde{m}_2\\
\tilde{m}_3\neq \tilde{m}_4
\end{aligned}
\right\}
\end{eqnarray}
\begin{eqnarray}
    \epsilon_{\tilde{m}_3}(\vk')= \epsilon_{\tilde{m}_4}(-\vk')\label{eqn:isingenergy}
\end{eqnarray}
and Eq.~(\ref{eqn:gapform}) can be rewritten as
\begin{eqnarray}
\label{eqn:gapformising}
 \Delta^{\tilde{m}_{1},\tilde{m}_{2}}_{}(\vk)&=&-\sum_{\substack{n, \tilde{m}_3 , \tilde{m}_4 ,  \\\vk' }}  \frac{V^{\tilde{m}_2,\tilde{m}_1}_{\tilde{m}_3,\tilde{m}_4}(\vk,\vk')\Delta^{\tilde{m}_{3},\tilde{m}_{4}}_{}(\vk')}{\omega^{2}_{n} + \epsilon^{2}_{\tilde{m}_3}(\vk')}.
\end{eqnarray}

The definition for gap function should have an extra $k_{\mathrm{B}}T$ factor that we ignore here (this factor cancels with a similar term from the Fourier transform of the Greens function so the form of gap function is not affected).

For the two band system considered here, the wave vectors take two sets of values for $\tilde{m}=(n,\sigma)$. These are $\tilde{m}=(1,\uparrow)$ for set of points $\vk_1$ on band 1 and $\tilde{m}=(2,\downarrow)$ for set of points $\vk_2$ on band 2. Note that for a given wavevector $\vk_1$ on band 1, we must have $-\vk_1$ lying on band 2 (i.e the vector $-\vk_1$ lies within the set of points $\vk_2$). From equation \ref{eqn:gapform} we split the momentum sum into sum over band 1 and band 2.

\begin{eqnarray}
 \Delta^{1,2}_{\uparrow\downarrow}(\vk_{1})&=&-\sum_{n\vk'_{{1}}}  \frac{V^{2,1,\downarrow,\uparrow}_{1,2,\uparrow,\downarrow}(\vk_{{1}},\vk_{{1}}')\Delta^{1,2}_{\uparrow \downarrow}(\vk_{{1}}')}{\omega^{2}_{n} + \epsilon^{2}_{1 \uparrow}(\vk_{{1}}')}
 -\sum_{n\vk'_{{2}}}  \frac{V^{2,1,\downarrow,\uparrow}_{2,1,\downarrow,\uparrow}(\vk_1,\vk_{{2}}')\Delta^{2,1}_{\downarrow \uparrow}(\vk_{{2}}')}{\omega^{2}_{n} + \epsilon^{2}_{2 \downarrow}(\vk_{{2}}')}
 \label{eqn:gap_12}
\end{eqnarray}

A similar equation is obtained for $\Delta^{2,1}_{\downarrow\uparrow}(\vk_{2})$. Note that the above equation considers zero momentum and zero frequency pairings only. Since the above equation explicitly incorporates the band index, it includes the points of degeneracy where the Ising spin orbit coupling vanishes leading to $\epsilon_{1\uparrow}(\vk)=\epsilon_{2\downarrow}(\vk)$. 

The above sum over the discrete Fermi surface points can be expressed in the form $\sum_{\vk_i} f(\vk_i)  \rightarrow \int d\epsilon_i\langle f(\vk_i) \rangle_{F.S}$. Note that the average over the Fermi surface containing discrete wavevectors $\vk_i$ can be expressed as $\langle f(\vk_i) \rangle_{F.S}=\frac{1}{N_k}\sum_{\vk_i}f(\vk_i)$. The linearized gap equation would then be a coupled set of equations in the variables $\vk_i$ and quantum numbers representing the superconducting gap function. After performing the energy integral and Matsubara frequency sums, these coupled equations can be expressed as a matrix equation given by,
\begin{eqnarray}
    \begin{pmatrix}
 \Delta^{1,2}_{\uparrow,\downarrow}(\vk_{F_{1}}) \\
 \Delta^{2,1}_{\downarrow,\uparrow}(\vk_{F_{2}})
\end{pmatrix}
=-\ln\left(\frac{2\omega_c e^{\gamma}}{\pi k_{\mathrm{B}}T_c}\right)
\begin{pmatrix}
 \frac{V^{2,1,\downarrow,\uparrow}_{1,2,\uparrow,\downarrow}(\vk_{F_{1}},\vk_{F_{1}}')}{|\nabla\epsilon_{1,\uparrow}|_{\vk'=\vk_{F_{1}}'}} & \frac{V^{2,1,\downarrow,\uparrow}_{2,1,\downarrow,\uparrow}(\vk_{F_{1}},\vk_{F_{2}}')}{|\nabla\epsilon_{2,\downarrow}|_{\vk'=\vk_{F_{2}}'}} \\
 \frac{V^{1,2,\uparrow,\downarrow}_{1,2,\uparrow,\downarrow}(\vk_{F_{2}},\vk_{F_{1}}')}{|\nabla\epsilon_{1,\uparrow}|_{\vk'=\vk_{F_{1}}'}} & \frac{V^{1,2,\uparrow,\downarrow}_{2,1,\downarrow,\uparrow}(\vk_{F_{2}},\vk_{F_{2}}')}{|\nabla\epsilon_{2,\downarrow}|_{\vk'=\vk_{F_{2}}'}} 
\end{pmatrix}
   \begin{pmatrix}
 \Delta^{1,2}_{\uparrow,\downarrow}(\vk_{F_{1}}') \\
 \Delta^{2,1}_{\downarrow,\uparrow}(\vk_{F_{2}}'),
\end{pmatrix}
\end{eqnarray}
where $\omega_c$ is some cutoff that we assume to be the same for both band integrals. Note that for N $\vk_{F_{1}}$ points, and N $\vk_{F_{2}}$ points, the dimension of the gap equation matrix is $2N \times 2N$\footnote{This form of the linearized gap equation (in absence of spin-orbit coupling) has been discussed widely for example in Refs.~\cite{Graser2009,wang2013superconducting,kemper2010sensitivity,wu2015} where the Fermi surface integral is discretized as well. However, this last step of writing the equation as matrix equation is often not given explicitly.}.  The above linearized superconducting gap equation can be solved to determine the ground state gap solution. If the set of points on the band 1 Fermi surface are $\vk_{F_{1}}=(\vk_{1F_{1}},\vk_{2F_{1}},\vk_{3F_{1}},\vk_{4F_{1}} \cdot \cdot \cdot \text{N points})$, and the N points on band 2 are ordered as $\vk_{F_{2}}=(-\vk_{1F_{1}},-\vk_{2F_{1}},-\vk_{3F_{1}},-\vk_{4F_{1}} \cdot \cdot \cdot \text{N points})$ then using Eq~\ref{eqn:vproperty}, the off diagonal blocks of the gap equation matrix have the symmetry $V^{1,2,\uparrow,\downarrow}_{2,1,\downarrow,\uparrow}(\vk_{F_{1}},\vk_{F_{2}}')=V^{2,1,\downarrow,\uparrow}_{1,2,\uparrow,\downarrow}(-\vk_{F_{1}},-\vk_{F_{2}}')=V^{2,1,\downarrow,\uparrow}_{1,2,\uparrow,\downarrow}(\vk_{F_{2}},\vk_{F_{1}}')$. Similarly the function $A_{\tilde{m}}(\vk)=1/|\nabla \epsilon_{\tilde{m}} (\vk)|$ on the two bands are related by $A_{1,\uparrow}(\vk_1)=A_{2,\downarrow}(-\vk_1)$

Below we describe the procedure for performing the Matsubara sum and energy integral. We express the R.H.S. of the gap equation (\ref{eqn:gap_12}) as
\begin{eqnarray}
    I_{\uparrow \downarrow}=\langle f_{\uparrow\downarrow}\rangle_{\text{F.S}} I(B).
\end{eqnarray}
with the function
\begin{eqnarray}
 I(B)=-\frac{1}{\beta} \sum_{n-\infty}^{\infty} \int \frac{1}{((i\omega_{n}-g\mu_{B}B)-\epsilon )((i\omega_{n}-g\mu_{B}B)+\epsilon )}  \,d\epsilon.
  \label{eqn:mat_first}
\end{eqnarray}
To perform the above integration  we will take help of complex integration and use residue theorem
\begin{eqnarray}
  \oint \frac{1}{ (z-(i\omega_{n}-g\mu_{B}B))(z+(i\omega_{n}-g\mu_{B}B)) }   \,dz.
\end{eqnarray}
Here, $\omega_{n}=(2n+1)\frac{\pi}{\beta}$ are the Fermionic Matsubara frequencies, and we perform the integration in the upper half plane. We have two simple poles at $z=(i \omega_{n}-g\mu_{B}B)$ with $n \geq 0$ and  $z=-(i \omega_{n}-g\mu_{B}B)$ with $n < 0$ respectively. Hence from the residue theorem
\begin{eqnarray}
 I(B)&=&-\frac{1}{\beta} \sum_{n-\infty}^{\infty} \int \frac{1}{((i\omega_{n}-g\mu_{B}B)-\epsilon )((i\omega_{n}-g\mu_{B}B)+\epsilon )}  \,d\epsilon \nonumber \\
 &=&\frac{2\pi i}{\beta}  \Bigl\{ \sum_{n \geq 0 }^{}\frac{1}{2(i \omega_{n}-g\mu_{B}B)}+\sum_{n<0}^{}\frac{-1}{2(i \omega_{n}-g\mu_{B}B)} \Bigr\}.
\end{eqnarray}
We can change the limit of $n$ in second term of R.H.S and rewrite above equation as
\begin{eqnarray}
 I(B) &=&\frac{\pi}{\beta}   \sum_{n \geq 0 }^{}\{\frac{1}{(| \omega_{n}|+i g\mu_{B}B)}+\frac{1}{(| \omega_{n}|-i\,g\mu_{B}B)} \} \nonumber \\
 &=& 2\mathrm{Re} \, \left[  \sum_{n \geq 0 }^{}\frac{1}{(| \omega_{n}|+i g\mu_{B}B)}\right].
\end{eqnarray}
This Matsubara sum with proper upper limit can be easily calculated:
\begin{eqnarray}
    \frac{\pi}{\beta} \sum_{n \geq 0 }^{} \frac{1}{|\omega_{n}|+x}=\ln(\frac{2 \omega_{c} e^{\gamma}\beta}{\pi})-F(\frac{k_{\mathrm{B}}T}{x}).
\end{eqnarray}
Here, $\gamma\approx 0.577$ is Euler's constant,
\begin{eqnarray}
  F(z)  =\psi(\frac{1}{2}+\frac{1}{2\,\pi\,z})-\psi(\frac{1}{2})
\end{eqnarray}
where $\psi$ denotes the digamma function.
Hence,
\begin{eqnarray}
 I(B) &=&2\ln (\frac{2 \omega_{c} e^{\gamma}\beta}{\pi})-2\mathrm{Re}\left[\psi(\frac{1}{2}+\frac{i g\mu_{B}B }{2\,\pi k_{\mathrm{B}}T_c})-\psi(\frac{1}{2})\right]\nonumber\\
 &=& 2 \ln(\frac{2 \omega_{c} e^{\gamma} \beta}{\pi})-2 \mathrm{Re}\left[F\left(\frac{k_{\mathrm{B}}T_c}{i g\mu_{B}B}\right)\right]
 \label{eqn:mat_final}
\end{eqnarray}
An identical term arises from the second part of equation (\ref{eqn:gap_12}). Since we are intregrating over the upper half plane, a factor of $(1/2)$ has to be multiplied with equation  (\ref{eqn:mat_final}).\\

\subsection{In Presence of Out of Plane Magnetic Field}

In the following we study the effect of a longitudinal magnetic field to the monolayer NbSe$_2$. Although near the transition temperature the orbital effect (vector potential contribution) dominates, in this section we ignore the orbital effect and consider only the Pauli paramagnetism contribution within the superconducting state. In the orbital basis, the Zeeman field Hamiltonian will be given by
\begin{eqnarray}
H_{B_z}=g\mu_B B_z \sum_{l,\vk}(n_{\vk l \uparrow}-n_{\vk l \downarrow}),
\end{eqnarray}
where $n_{\vk l \sigma}=c^{\dag}_{\vk l \sigma}c_{\vk l \sigma}$. Assuming the non interacting Hamiltonian has no spin flip contribution, the above term keeps $S_z$ as a good quantum number. If we diagonalise the total Hamiltonian with the Zeeman term included, we can look at a model that considers the modification to the pairing term due to the magnetic field using a similar approach to the zero field case discussed above. We will only need to modify the gap equation in Eq~\ref{eqn:gapform}. As discussed above the gap equation is a coupled set of equations  
\begin{eqnarray}
 \Delta^{\tilde{m}_{1},\tilde{m}_{2}}_{}(\vk)&=&\sum_{\substack{n, \tilde{m}_3 , \tilde{m}_4 ,\\\vk' }}  \frac{V^{\tilde{m}_2,\tilde{m}_1}_{\tilde{m}_3,\tilde{m}_4}(\vk,\vk',B_z)\Delta^{\tilde{m}_{3},\tilde{m}_{4}}_{}(\vk')}{(i\omega_{n} - \epsilon_{\tilde{m}_3}(\vk'))(i\omega_{n} + \epsilon_{\tilde{m}_4}(-\vk')) } 
\end{eqnarray}
Here, the magnetic field dependence enters both the pairing interaction and the single particle energies.
By breaking the Kramers degeneracy the modified relation between the single particle energy on the two bands relevant to monolayer NbSe$_2$ read $\epsilon_{m_3/m_4}\rightarrow \epsilon_{m_3/m_4} \pm g\mu_B B_z$ since the indices $m_3$ and $m_4$ on the Fermi surface necessarily belong to the spin up and spin down bands or vice versa. This is also true for the Hamiltonian that ignores the spin orbit contribution when restricted to only opposite spin pairing singlet or triplet solutions. This implies the gap equation modifies to
\begin{eqnarray}
 \Delta^{\tilde{m}_{1},\tilde{m}_{2}}_{}(\vk)=-\sum_{\substack{n, \tilde{m}_3 , \tilde{m}_4 ,\\\vk' }}  \frac{V^{\tilde{m}_2,\tilde{m}_1}_{\tilde{m}_3,\tilde{m}_4}(\vk,\vk',B_z)\Delta^{\tilde{m}_{3},\tilde{m}_{4}}(\vk')}{\tilde{\omega}_{n}^2 + \epsilon^2_{\tilde{m}_3}(\vk') }
\end{eqnarray}
where the Matsubara frequency is shifted by the field, $\tilde{\omega}_n=\omega_n \pm ig\mu_B B$, and the sign $\pm$ corresponding to $(\tilde{m}_{3},\tilde{m}_{4})=(2\downarrow,1\uparrow)$ or $(\tilde{m}_{4},\tilde{m}_{3})=(1\uparrow,2\downarrow)$ respectively. Note that the symmetry  $\epsilon^2_{\tilde{m}_3}(\vk')=\epsilon_{\tilde{m}_4}(-\vk')$, is generically true for Ising SOC on any constant energy surface. For no spin orbit coupling we have the additional criterion $\tilde{m}_1=\tilde{m}_2$, and $\tilde{m}_3=\tilde{m}_4$ if we assume zero energy and zero momentum pairing.

For weak magnetic fields, we can consider the magnetic field to be a perturbation to the non interacting Hamiltonian. In this limit, we consider the Fermi surface to be the one obtained in zero magnetic field. Note that this Zeeman term that acts as a perturbation, will still make the energy integral and correspondingly the transition temperature field dependant but may not modify the gap matrix. It is interesting to ask how this formalism would differentiate between the even and odd parity superconducting states. Let us re-write the gap equation relevant to monolayer NbSe$_2$ in the presence of Ising SOC.
\begin{eqnarray}
\Delta^{1,2}(\vk_1)=-\sum_{\substack{n,\vk'_1 }}  \frac{V^{2,1}_{1,2}(\vk_1,\vk'_1)\Delta^{1,2}(\vk'_1)}{\tilde{\omega}_{n+}^2 + \epsilon^2_{1}(\vk'_1) } - \sum_{\substack{n,\vk'_2 }}  \frac{V^{2,1}_{2,1}(\vk_2,\vk'_2)\Delta^{2,1}(\vk'_2)}{\tilde{\omega}_{n-}^2 + \epsilon^2_{2}(\vk'_2) }\\
\Delta^{2,1}(\vk_2)=-\sum_{\substack{n,\vk'_1 }}  \frac{V^{1,2}_{1,2}(\vk_2,\vk'_1)\Delta^{1,2}(\vk'_1)}{(\tilde{\omega}_{n+}^2 + \epsilon^2_{1}(\vk'_1)) } - \sum_{\substack{n,\vk'_2 }}  \frac{V^{1,2}_{2,1}(\vk_2,\vk'_2)\Delta^{2,1}(\vk'_2)}{\tilde{\omega}_{n-}^2 + \epsilon^2_{2}(\vk'_2) }
\end{eqnarray}
where $\tilde{\omega}_{n\pm}=\omega_n \pm ig\mu_B B$. As discussed before, for Ising spin orbit coupling, the band indices are attached to the corresponding spin indices $(1,\uparrow)$, and $(2,\downarrow)$. The gap equation results in a magnetic field dependence in the transition temperature
\begin{eqnarray}
    \begin{pmatrix}
 \Delta^{1,2}_{\uparrow,\downarrow}(\vk_{F_{1}}) \\
 \Delta^{2,1}_{\downarrow,\uparrow}(\vk_{F_{2}})
\end{pmatrix}
=-\left[\ln\left(\frac{2\omega_c e^{\gamma}}{\pi k_{\mathrm{B}}T_c}\right)- \mathop{\mathrm{Re}} \left[ F\left( \frac{k_{\mathrm{B}}T_c}{i g\mu_{B}B} \right) \right] \right]
\begin{pmatrix}
 \frac{V^{2,1,\downarrow,\uparrow}_{1,2,\uparrow,\downarrow}(\vk_{F_{1}},\vk_{F_{1}}')}{|\nabla\epsilon_{1,\uparrow}|_{\vk'=\vk_{F_{1}}'}} & \frac{V^{2,1,\downarrow,\uparrow}_{2,1,\downarrow,\uparrow}(\vk_{F_{1}},\vk_{F_{2}}')}{|\nabla\epsilon_{2,\downarrow}|_{\vk'=\vk_{F_{2}}'}} \\
 \frac{V^{1,2,\uparrow,\downarrow}_{1,2,\uparrow,\downarrow}(\vk_{F_{2}},\vk_{F_{1}}')}{|\nabla\epsilon_{1,\uparrow}|_{\vk'=\vk_{F_{1}}'}} & \frac{V^{1,2,\uparrow,\downarrow}_{2,1,\downarrow,\uparrow}(\vk_{F_{2}},\vk_{F_{2}}')}{|\nabla\epsilon_{2,\downarrow}|_{\vk'=\vk_{F_{2}}'}} 
\end{pmatrix}
   \begin{pmatrix}
 \Delta^{1,2}_{\uparrow,\downarrow}(\vk_{F_{1}}') \\
 \Delta^{2,1}_{\downarrow,\uparrow}(\vk_{F_{2}}')
\end{pmatrix}
\label{eqn_out_of_plane_magnetic_field_effect}
\end{eqnarray}

Here, $ F(z)  =\psi(\frac{1}{2}+\frac{1}{2\,\pi\,z})-\psi(\frac{1}{2})$, $\psi$ being  digamma function. As can be seen from Eq(\ref{eqn_out_of_plane_magnetic_field_effect}), the gap structure is not affected by the presence of a magnetic field. In particular, since the magnetic field does not explicitly enter the gap matrix itself, the Pauli paramagnetic effect for magnetic fields perpendicular to the plane of NbSe$_2$, is similar for spin singlet, and spin triplet states with opposite spin pairing.\\

\subsection{In Presence of In-Plane Magnetic Field}
The modified Gorkov equations for a translationally invariant system (and restricting ourselves to homogeneous solutions), the magnetic field would introduce an additional term associated with $h$, 
\begin{eqnarray}
(i \omega_{n} - \epsilon_{m\sigma}(\vk)) G^{m,m'}_{\sigma,\sigma'}(\vk,\omega_{n}) 
-\sum_{s} h_{\vk \sigma \bar{\sigma}}^{m,s}G^{s,m'}_{\bar{\sigma},\sigma'}(\vk,\omega_{n})+\sum_{m_1,\sigma_{1} }  \Delta^{m,m_{1}}_{\sigma,\sigma_{1}}(\vk) F^{\ast m_{1},m'}_{\sigma_{1},\sigma'}(\vk,\omega_{n}) &=&\delta_{m,m'}\delta_{\sigma,\sigma'} \label{eqn:gorkovbx1}\\
(i\omega_{n} - \epsilon_{m\sigma}(\vk)) F^{m,m'}_{\sigma,\sigma'}(\vk,\omega_{n})
- \sum_s h_{\vk \sigma \bar{\sigma}}^{m,s} F^{s,m'}_{\bar{\sigma},\sigma'}(\vk,\omega_{n}) 
-\sum_{m_1\sigma_{1}}  \Delta^{m,m_{1}}_{\sigma,\sigma''}(\vk) G^{m',m_1}_{\sigma',\sigma_{1}}(-\vk,-\omega_{n}) &=&0\label{eqn:gorkovbx2}
\end{eqnarray}
%
Before discussing the gap equation in the presence of an Ising spin orbit coupling, let us first look at the gap equation in the absence of an Ising spin orbit coupling. From the above Gorkov equations, the gap equation in the absence of SO coupling can be expressed as
\begin{eqnarray}
\Delta^{mm}_{\uparrow \downarrow}(\vk)=-2\sum_{n\vk'm'}\frac{V^{\downarrow \uparrow  mm}_{\uparrow \downarrow m'm'}(\vk,\vk')}{Z^{m'm'}_{}(\vk,\omega_n)Z^{m'm'}_{}(-\vk,-\omega_n)}\left((\omega_n^2+\epsilon_{m'\vk}^2)\Delta^{m'm'}_{\uparrow \downarrow}(\vk')+h^2\Delta^{m'm'}_{\downarrow \uparrow}(\vk')\right)
\end{eqnarray}
Here, $Z^{m,m}(\vk,\omega_n)=((i \omega_{n} - \epsilon_{m\sigma}(\vk)(i \omega_{n} - \epsilon_{m\bar{\sigma}}(\vk))-h^2$, and we also use the property $\epsilon_{m,\sigma}(\vk)=\epsilon_{m,\bar{\sigma}}(\vk)=\epsilon_{m,\sigma}(-\vk)=\epsilon_{m,\sigma}(\vk)$. Similar equations can be written for the $\Delta^{mm}_{\downarrow \uparrow}(\vk)$ gap function. 

Notice that unlike the case for c-axis magnetic field the gap equation for the spin singlet state and opposite pairing spin triplet state are not equivalent for in plane magnetic fields even in the absence of spin orbit coupling.
Let us now consider a finite Ising SOC. In the above equations we have $\bar{\sigma}=-\sigma$ and $h_{\vk \sigma \bar{\sigma}}^{m,s}=g\mu_BB_x\sum_{\mu}u_{\mu \sigma}^{*m}(\vk)u_{\mu \bar{\sigma}}^{s}(\vk)$, where $u$ is the unitary transformation from orbital to band space. $h$ satisfy following relations
\begin{eqnarray}
h_{\vk \bar{\sigma }\sigma}^{m',m}&=&g\mu_BB_x\sum_{\mu}u_{\mu \bar{\sigma}}^{*m'}(\vk)u_{\mu\sigma }^{m}(\vk)=h_{\vk\sigma \bar{\sigma }}^{*m,m'}\\
h_{-\vk \bar{\sigma }\sigma}^{m',m}&=&g\mu_BB_x\sum_{\mu}u_{\mu \bar{\sigma}}^{*m'}(-\vk)u_{\mu\sigma }^{m}(-\vk)=g\mu_BB_x\sum_{\mu}u_{\mu \sigma}^{*m}(\vk)u_{\mu\bar{\sigma}}^{m'}(\vk)=h_{\vk\sigma \bar{\sigma }}^{m,m'}
\end{eqnarray}
The definition of the gap function is as in Eq~\ref{eqn:deltaform}
\begin{eqnarray}
\label{eqn:deltaformbx}
 \Delta^{m_{1},m_{2}}_{\sigma_1,\sigma_2}(\vk)&=&-\sum_{\substack{n, m_3 , m_4 ,  \\\vk }}  V^{m_2,m_1,\sigma_2,\sigma_1}_{m_3,m_4,\sigma_3,\sigma_4}(\vk,\vk')F^{m_{3},m_{4}}_{\sigma_3,\sigma_4}(\vk',\omega_{n})
\end{eqnarray}
where the pairing potential is computed in absence of external magnetic field. Note that due to Ising SOC, the band and spin indices are coupled in pairs $(m,\sigma)$.

The presence of the in-plane Zeeman field allows for opposite spin correlations in the particle hole Greens functions and equal spin correlations in the particle particle Green's function. For a generic multi band system the above equations lead to a complicated solution. However, if we restrict our analysis to monolayer NbSe$_2$, then the indices take two possible values $(m,\sigma)=(1,\uparrow)$ or $(2,\downarrow)$ at low energies. The above equations then take the form
\begin{eqnarray}
(i \omega_{n} - \epsilon_{m\sigma}(\vk)) G^{m,m'}_{\sigma,\sigma'}(\vk,\omega_{n}) 
-h_{\vk \sigma \bar{\sigma}}^{m,s}G^{s,m'}_{\bar{\sigma},\sigma'}(\vk,\omega_{n})+ \sum_{m_1,\sigma_1}\Delta^{m,m_{1}}_{\sigma,\sigma_1}(\vk) F^{\ast m_{1},m'}_{\sigma_1,\sigma'}(\vk,\omega_{n}) &=&\delta_{mm'}\delta_{\sigma\sigma'} \label{eqn:gorkovbxnb1}\\
(i\omega_{n} - \epsilon_{m\sigma}(\vk)) F^{m,m'}_{\sigma,\sigma'}(\vk,\omega_{n})
-  h_{\vk \sigma \bar{\sigma}}^{m,s} F^{s,m'}_{\bar{\sigma},\sigma'}(\vk,\omega_{n}) 
-\sum_{m_1,\sigma_1}  \Delta^{m,m_{1}}_{\sigma,\sigma_1}(\vk) G^{m',m_1}_{\sigma',\sigma_1}(-\vk,-\omega_{n}) &=&0\label{eqn:gorkovbxnb2}
\end{eqnarray}
If we now consider the linearised equations then from Eq~\ref{eqn:gorkovbxnb1}  we get
\begin{eqnarray}
G^{(0)m',m}_{\bar{\sigma},\sigma}(\vk,\omega_{n} ) &=& \frac{ \, h_{\vk ,\bar{\sigma} \sigma}^{m',m}G^{m,m}_{\sigma,\sigma}(\vk,\omega_{n})}{i \omega_{n} - \epsilon_{m'\bar{\sigma}}(\vk)
}\label{eqn:l_85}\\
G^{(0)m,m}_{\sigma,\sigma}(\vk,\omega_{n})&=& \frac{i \omega_{n} - \epsilon_{m'\bar{\sigma}}(\vk)}{(i \omega_{n} - \epsilon_{m'\bar{\sigma}}(\vk))(i \omega_{n} - \epsilon_{m\sigma}(\vk))-|h_{\vk \sigma \bar{\sigma}}^{m,m'}|^2}
\label{eqn:l_86}
\end{eqnarray}

Solving Eq~\ref{eqn:gorkovbxnb2} with aid of Eq~\ref{eqn:l_85} and Eq~\ref{eqn:l_86} we get
\begin{eqnarray}
F_{\sigma \sigma}^{mm}(\vk,\omega_n)=\frac{h_{\vk \sigma \bar{\sigma}}^{mm'}F_{\bar{\sigma}\sigma}^{m'm}(\vk,\omega_n)}{i\omega_n-\epsilon_{m\sigma}(\vk)}+\frac{h_{-\vk \sigma \bar{\sigma}}^{mm'}\Delta_{\sigma \bar{\sigma}}^{mm'}(\vk)G_{\bar{\sigma}\bar{\sigma}}^{m'm'}(-\vk,-\omega_n)}{(i\omega_n-\epsilon_{m\sigma}(\vk))(-i\omega_n-\epsilon_{m\sigma}(-\vk))}
\end{eqnarray}
As expected the equal spin pairing correlations are induced by the in plane magnetic field. The opposite spin correlation is given by
\begin{eqnarray}
F_{\sigma \bar{\sigma}}^{mm'}(\vk,\omega_n)&=&-\frac{\Delta_{\sigma \bar{\sigma}}^{mm'}(\vk)(i\omega_n-\epsilon_{m'\bar{\sigma}}(\vk))(i\omega_n+\epsilon_{m\sigma}(-\vk))}{Z_{\sigma\bar{\sigma}}^{m,m'}(\vk,i \omega_{n})Z_{\bar{\sigma}\sigma }^{m',m}(-\vk,-i \omega_{n})}+\frac{|h_{\vk \sigma \bar{\sigma}}^{mm'}|^2\Delta_{\bar{\sigma} \sigma}^{m'm}(\vk)}{Z_{\sigma\bar{\sigma}}^{m,m'}(\vk,i \omega_{n})Z_{\bar{\sigma}\sigma }^{m',m}(-\vk,-i \omega_{n})} \nonumber\\
\end{eqnarray}
Here, $Z_{\sigma\bar{\sigma}}^{m,m'}(\vk,i \omega_{n})=(i \omega_{n} - \epsilon_{m\sigma}(\vk))(i \omega_{n} - \epsilon_{m'\bar{\sigma}}(\vk))-|h_{\vk \sigma \bar{\sigma}}^{m,m'}|^2 $

The in-plane field induces an additional component in the anomalous Green functions $F$. This implies that the superconducting gap equation will be modified by the additional in plane field contribution.

The above function would lead to a complicated superconducting gap equation where the energy integral cannot be solved independently because of the momentum dependence of the magnetic field.\\
Additionally notice that in the absence of the magnetic field the energy dependence in the linearized gap equation always entered the equation as a product of the form $\epsilon_{m\sigma}(\vk)\epsilon_{m'\bar{\sigma}}(-\vk)$. Then for both cases of no Ising SOC (inversion symmetric) ($\epsilon_{m\sigma}(\vk)=\epsilon_{m\sigma}(-\vk)=\epsilon_{m'\bar{\sigma}}(\vk)$) or presence of Ising SOC ($\epsilon_{m\sigma}(\vk)=\epsilon_{m'\bar{\sigma}}(-\vk)$) the gap equation could be written in terms of an energy integral over a single band index. However in the presence of an in plane magnetic field the gap equation contains product terms of the form $\epsilon_{m\sigma}(\vk)\epsilon_{m'\bar{\sigma}}(\vk)$. Such terms cannot be converted straightforwardly into the energy integral over a single band index.  We therefore express the property of Ising, $\epsilon_{m\sigma}(\vk)=\epsilon_{m'\bar{\sigma}}(-\vk)$ and define $2\delta_{\vk}$ = $\epsilon_{m'\bar{\sigma}}(\vk)-\epsilon_{m\sigma}(\vk)$ (where $\delta_{\vk}$ is known to us numerically and is non zero in the presence of Ising SOC. We are working with the assumption that at Fermi energy there is a single band that splits into two relevant to NbSe$_2$.). We then get the following equation

\begin{eqnarray}
F_{\sigma \bar{\sigma}}^{mm'}(\vk,\omega_n)&=&-\frac{\Delta_{\sigma \bar{\sigma}}^{mm'}(\vk)(i\omega_n-\left(2\delta_{\vk}+\epsilon_{m\sigma}(\vk)\right))(i\omega_n+\left(2\delta_{\vk}+\epsilon_{m\sigma}(\vk)\right))}{Z_{\sigma\bar{\sigma}}^{m,m'}(\vk,i \omega_{n})Z_{\bar{\sigma}\sigma }^{m',m}(-\vk,-i \omega_{n})}+\frac{|h_{\vk \sigma \bar{\sigma}}^{m m'}|^2\Delta_{\bar{\sigma} \sigma}^{m'm}(\vk)}{Z_{\sigma\bar{\sigma}}^{m,m'}(\vk,i \omega_{n})Z_{\bar{\sigma} \sigma }^{m',m}(-\vk,-i \omega_{n})}\nonumber\\\label{eqn:l_89}
\end{eqnarray}

Here, $Z_{\sigma\bar{\sigma}}^{m,m'}(\vk,i \omega_{n})$ gets modified as $Z_{\sigma\bar{\sigma}}^{m,m'}(\vk,i \omega_{n})=\left((i \omega_{n} - \epsilon_{m\sigma}(\vk))(i \omega_{n} - (2\delta_{k}+\epsilon_{m\sigma}(\vk)))-|h_{\vk \sigma \bar{\sigma}}^{m,m'}|^{2}\right) $. 

Now, the gap equation reads as

\begin{eqnarray}
 \Delta^{m_{1},m_{2}}_{\sigma,\bar{\sigma}}(\vk)=-\sum_{\substack{n, m_3 , m_4 \\\vk' }} \left[ V^{m_2,m_1,\bar{\sigma},\sigma}_{m_3,m_4,\sigma,\bar{\sigma}}(\vk,\vk')F^{m_{3},m_{4}}_{\sigma,\bar{\sigma}}(\vk',\omega_{n})+ V^{m_2,m_1,\bar{\sigma},\sigma}_{m_3,m_4,\bar{\sigma},\sigma}(\vk,\vk')F^{m_{3},m_{4}}_{\bar{\sigma},\sigma}(\vk',\omega_{n})\right]
 \end{eqnarray}
Considering only opposite spin pairing with Ising Spin Orbit Coupling in mind we get

\begin{footnotesize}

\begin{eqnarray}
 -\Delta^{m_{1},m_{2}}_{\sigma,\bar{\sigma}}(\vk)
 &=&\sum_{\substack{n, m_3 , m_4 \\\vk' }}  \Delta_{\sigma \bar{\sigma}}^{mm'}(\vk)[V^{m_2,m_1,\bar{\sigma},\sigma}_{m_3,m_4,\sigma,\bar{\sigma}}(\vk,\vk')\frac{[(\epsilon_{m,\sigma}-i\omega_n)(\epsilon_{m,\sigma}+i\omega_n)+4\delta_{k}\epsilon_{m,\sigma}+4\delta_{k}^{2} ]}{Z_{\sigma\bar{\sigma}}^{m,m'}(\vk,i \omega_{n})Z_{\bar{\sigma}\sigma }^{m',m}(-\vk,-i \omega_{n})}\nonumber\\
 &+&V^{m_2,m_1,\bar{\sigma},\sigma}_{m_3,m_4,\bar{\sigma},\sigma}(\vk,\vk')\frac{ |h_{\vk  \bar{\sigma}\sigma}^{m'm}|^2}{Z_{\bar{\sigma}\sigma}^{m',m}(\vk,i \omega_{n})Z_{\sigma\bar{\sigma} }^{m,m'}(-\vk,-i \omega_{n})}]\nonumber\\
 &+& \Delta_{\bar{\sigma}\sigma }^{m'm}(\vk)[V^{m_2,m_1,\bar{\sigma},\sigma}_{m_3,m_4,\bar{\sigma},\sigma}(\vk,\vk')\frac{[(\epsilon_{m',\bar{\sigma}}-i\omega_n)(\epsilon_{m',\bar{\sigma}}+i\omega_n)-4\delta_{k}\epsilon_{m',\bar{\sigma}}+4\delta_{k}^{2} ]}{Z_{\bar{\sigma}\sigma }^{m',m}(\vk,i \omega_{n})Z_{\sigma\bar{\sigma} }^{m,m'}(-\vk,-i \omega_{n})}\nonumber\\
 &+&V^{m_2,m_1,\bar{\sigma},\sigma}_{m_3,m_4,\sigma,\bar{\sigma}}(\vk,\vk')\frac{ |h_{\vk  \sigma\bar{\sigma}}^{mm'}|^2}{Z_{\sigma\bar{\sigma}}^{m,m'}(\vk,i \omega_{n})Z_{\bar{\sigma}\sigma }^{m',m}(-\vk,-i \omega_{n})}]\nonumber\\
 \label{eqn:gap_couple}
 \end{eqnarray}
\end{footnotesize}
Therefore, the coupled BCS gap equation is of the form equation \ref{eqn:gap_couple}. Performing the energy integral similar as outlined in equations \ref{eqn:mat_first} to \ref{eqn:mat_final}, we can express the coupled equations in a matrix form as,
\begin{eqnarray}
    \begin{bmatrix}
 \Delta^{1,2}_{\uparrow,\downarrow}(\vk_{F_{1}}) \\
 \Delta^{2,1}_{\downarrow,\uparrow}(\vk_{F_{2}})
\end{bmatrix}
&=&-\sum_{\substack{\vk' }}
\begin{bmatrix}
 \left(\frac{I_{1}(\vk_{F_{1}}')V^{2,1,\downarrow,\uparrow}_{1,2,\uparrow,\downarrow}(\vk_{F_{1}},\vk_{F_{1}}')}{|\nabla\epsilon_{1,\uparrow}|_{\vk'=\vk_{F_{1}}'}}+
 \frac{I_{2}(\vk_{F_{2}}')V^{2,1,\downarrow,\uparrow}_{2,1,\downarrow,\uparrow}(\vk_{F_{1}},\vk_{F_{2}}')}{|\nabla\epsilon_{2,\downarrow}|_{\vk'=\vk_{F_{2}}'}}\right)
 & \left( \frac{I_{1}(\vk_{F_{2}}')V^{2,1,\downarrow,\uparrow}_{2,1,\downarrow,\uparrow}(\vk_{F_{1}},\vk_{F_{2}}')}{|\nabla\epsilon_{2,\downarrow}|_{\vk'=\vk_{F_{2}}'}}+\frac{I_{2}(\vk_{F_{1}}')V^{2,1,\downarrow,\uparrow}_{1,2,\uparrow,\downarrow}(\vk_{F_{1}},\vk_{F_{1}}')}{|\nabla\epsilon_{1,\uparrow}|_{\vk'=\vk_{F_{1}}'}}\right)
 \\
\left( \frac{I_{1}(\vk_{F_{1}}')V^{1,2,\uparrow,\downarrow}_{1,2,\uparrow,\downarrow}(\vk_{F_{2}},\vk_{F_{1}}')}{|\nabla\epsilon_{1,\uparrow}|_{\vk'=\vk_{F_{1}}'}}+
  \frac{I_{2}(\vk_{F_{2}}')V^{1,2,\uparrow,\downarrow}_{2,1,\downarrow,\uparrow}(\vk_{F_{2}},\vk_{F_{2}}')}{|\nabla\epsilon_{2,\downarrow}|_{\vk'=\vk_{F_{2}}'}}\right)
  & 
  \left(\frac{I_{1}(k_{F_{2}}')V^{1,2,\uparrow,\downarrow}_{2,1,\downarrow,\uparrow}(\vk_{F_{2}},\vk_{F_{2}}')}{|\nabla\epsilon_{2,\downarrow}|_{\vk'=\vk_{F_{2}}'}} +\frac{I_{2}(\vk_{F_{1}}')V^{1,2,\uparrow,\downarrow}_{1,2,\uparrow,\downarrow}(\vk_{F_{2}},\vk_{F_{1}}')}{|\nabla\epsilon_{1,\uparrow}|_{\vk'=\vk_{F_{1}}'}} \right)
  \label{eq:e_1}
\end{bmatrix}
\nonumber\\
&&\begin{bmatrix}
 \Delta^{1,2}_{\uparrow,\downarrow}(\vk_{F_{1}}') \\
 \Delta^{2,1}_{\downarrow,\uparrow}(\vk_{F_{2}}')
\end{bmatrix}
\end{eqnarray}
\begin{eqnarray}
    I_{1}(\vk)
    &=&- \text{ln}\left(\frac{T}{T_{\mathrm{c}}}\right)+\frac{1}{\lambda}- \frac{1}{2}\left(1-\frac{\delta_{\vk}^{2}}{\tilde{h}_{\vk}^{2}}\right)\left[\mathop{\mathrm{Re}}[\psi(\frac{1}{2}+i\frac{\tilde{h}_{\vk}}{2\pi k_{\mathrm{B}}T_{}})] -\psi(\frac{1}{2})\right]\\
      I_{2}(\vk)&=&\frac{1}{2}\left(1-\frac{\delta_{\vk}^{2}}{\tilde{h}_{\vk}^{2}}\right)\left[\mathop{\mathrm{Re}}[\psi(\frac{1}{2}+i\frac{\tilde{h}_{\vk}}{2\pi k_{\mathrm{B}}T_{}})] -\psi(\frac{1}{2})\right]
\end{eqnarray}

$[h_{k}^{2}+\delta_{\vk}^{2}]=\tilde{h}_{k}^{2}$

\section{Discussion on nesting vectors and Pairing structure}
\begin{figure}[h]
    \includegraphics[width=1.0\linewidth]{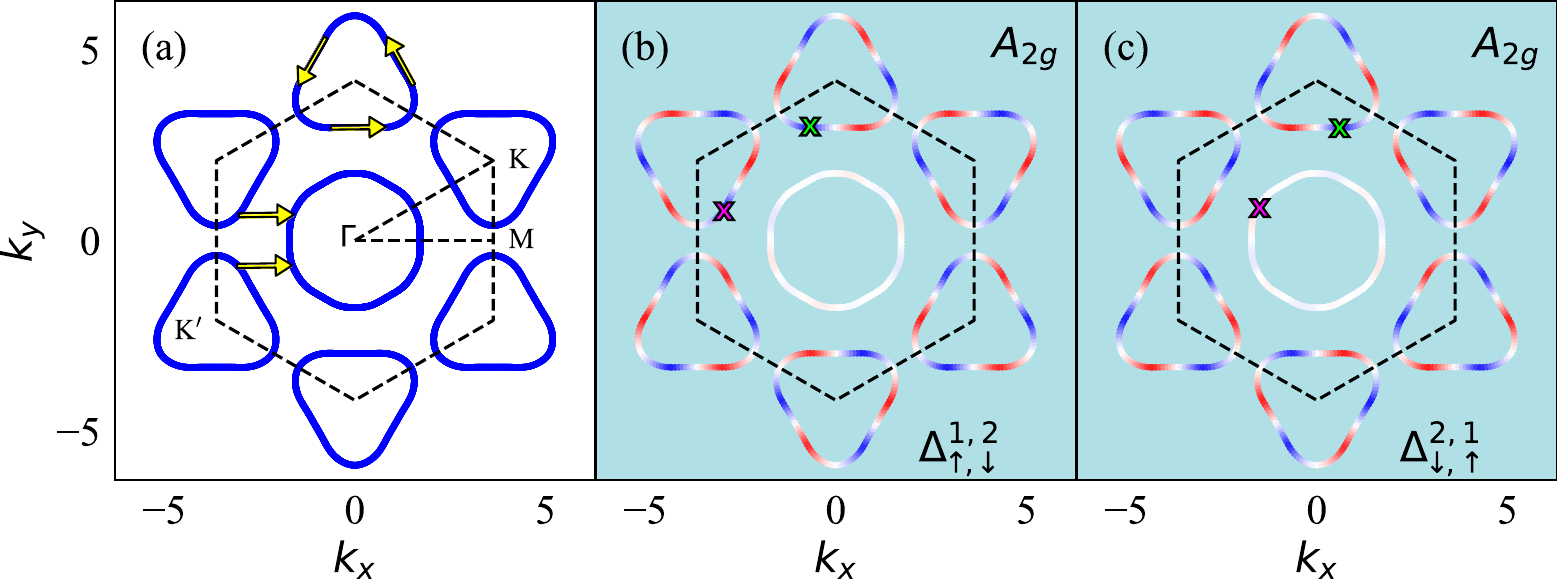}     
    \caption{(a) Fermi surface in absence of Ising SOC. Yellow arrows represent the nesting vectors. (b,c) Leading solution of the linearized gap equation
   ($\Delta^{1,2}_{\uparrow,\downarrow}(\vk_F)$ , $\Delta_{2,1}^{\downarrow,\uparrow}(\vk_F)$)
    plotted on the Fermi surface in absence of Ising SOC. Additional parameters: $U = 1.0$, $J = U/4$. Cross represent nesting vectors connecting gap on Fermi surface.}
    \label{fig_nesting_with_out}
\end{figure}
 The observed dominance of the nodal pairing structure can be attributed to the nesting properties of the Fermi surface. As shown in Figure 2(c) of the main text, a pronounced peak in susceptibility is observed around 0.4$\Gamma$M. This momentum vector aligns closely with the intra-pocket nesting vectors depicted by the yellow arrows in Fig~\ref{fig_nesting_with_out}(a). While inter-pocket nesting vectors connecting the $K$ and $\Gamma$ pockets exist, their relative scarcity suggests a negligible contribution. These dominant intra-pocket nesting vectors favor a strong pairing interaction within the $K$ pocket, leading to the observed nodal pairing structure, while the $\Gamma$-centered pocket exhibits a weaker pairing strength.
 
 \begin{figure}
    \includegraphics[width=1.0\linewidth]{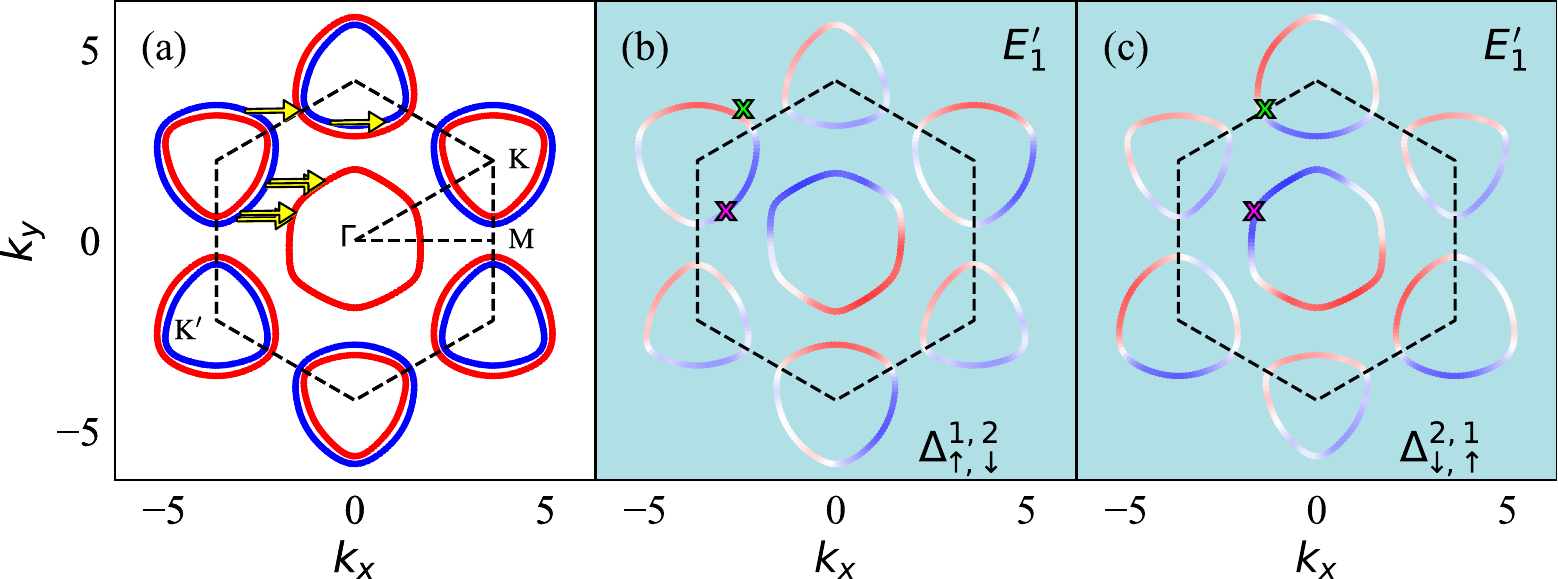}     
    \caption{(a) Spin split Fermi surface in presence of Ising SOC. For clarity, the $\Gamma$-centered pocket associated with up spin (blue) is not shown here.. Yellow arrows represent the nesting vectors. (b,c) Leading solution of the linearized gap equation
   ($\Delta^{1,2}_{\uparrow,\downarrow}(\vk_F)$ , $\Delta_{2,1}^{\downarrow,\uparrow}(\vk_F)$)
    plotted on the Fermi surface in presence of Ising SOC.
    Additional parameters: $U = 1.0$, $J = U/4$. Cross represent nesting vectors connecting gap on Fermi surface.}
    \label{fig_nesting_with_soc}
\end{figure}
Like before a pronounced peak in susceptibility is observed around 0.4$\Gamma$M, as shown in Figure 2(d) of the main text. However, due to the spin-splitting of the Fermi surface induced by Ising SOC, the $K$ and $\Gamma$ pockets exhibit a more pronounced curvature. This enhanced curvature facilitates the emergence of additional inter-pocket nesting vectors connecting these pockets. Consequently, in contrast to the scenario without Ising SOC, a significant pairing interaction now develops within the $\Gamma$-centered pocket.
\begin{figure}
    \includegraphics[width=1.0\linewidth]{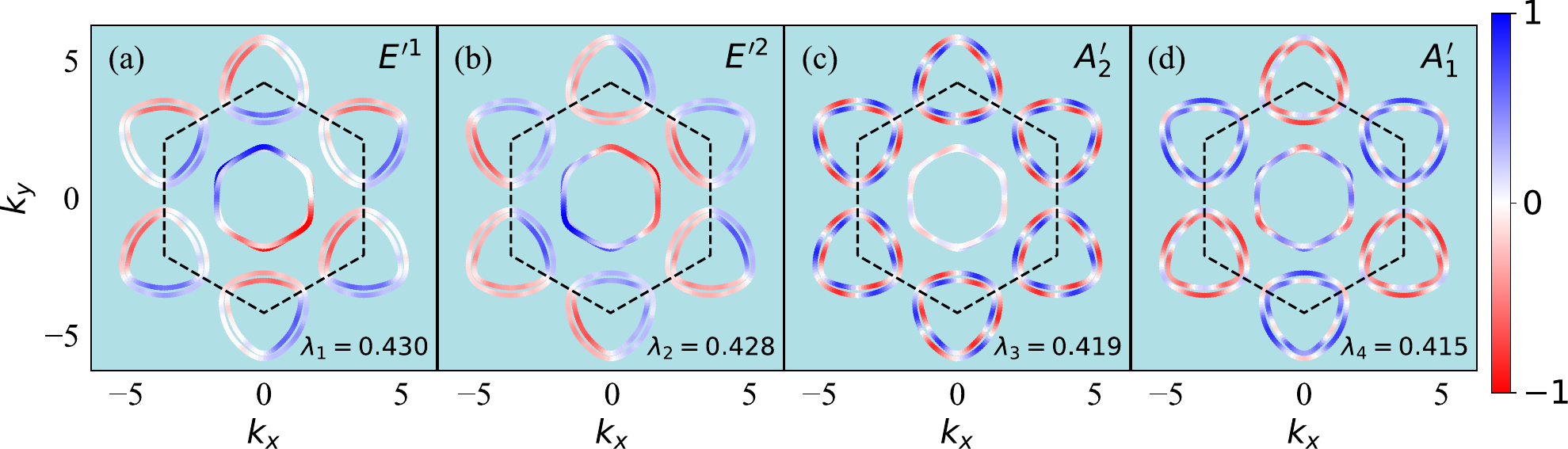}     
    \caption{Solutions of the linearized gap equation, $\Delta(\vk_F)$,
    plotted on the spin-split Fermi surface(D$_{3h}$ symmetry) for the four leading superconducting pairing gaps with corresponding eigenvalue $\lambda$. The non-interacting Hamiltonian used here neglects the Ising SOC and assumes D$_{6h}$ symmetry. Additional parameters: $U = 1.0$, $J = U/4$.}
    \label{fig_nesting_gap_with_out_soc}
\end{figure}
Figure~\ref{fig_nesting_gap_with_out_soc} presents the solutions of the linearized gap equation for a enforced spin-split Fermi surface with D$_{3h}$ symmetry and a pairing interaction calculated from a non-interacting Hamiltonian that neglects Ising SOC, thereby preserving its inherent D$_{6h}$ symmetry.  These results demonstrate that the spin-splitting of the Fermi points induced by Ising SOC is sufficient to drive the system towards a dominant spin-triplet pairing instability.
\begin{figure}
    \includegraphics[width=1.0\linewidth]{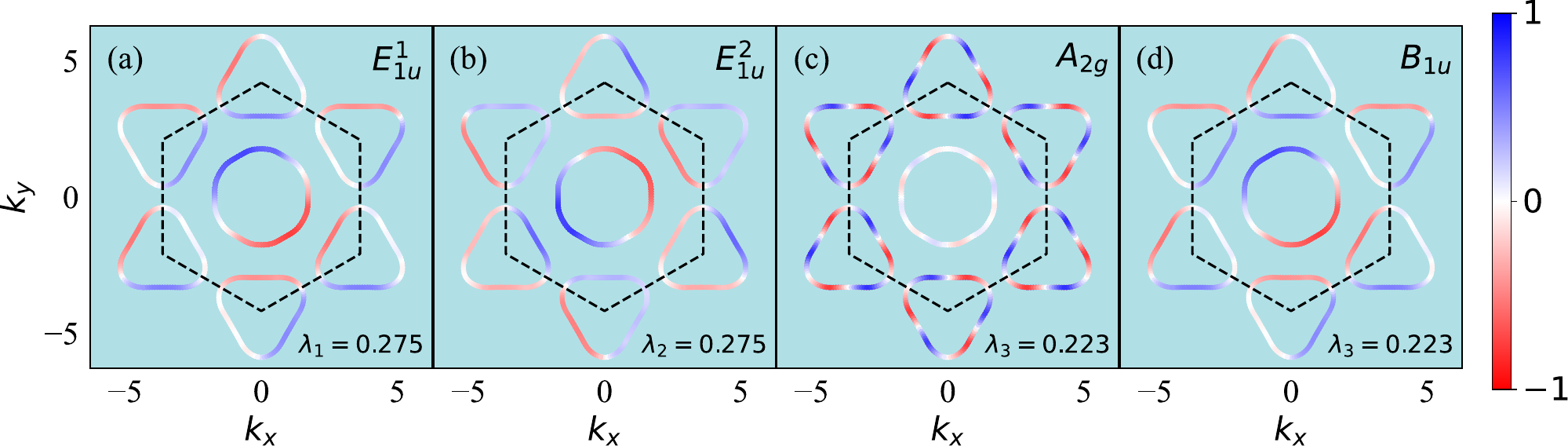}     
    \caption{Solutions of the linearized gap equation, $\Delta(\vk_F)$, plotted on the Fermi surface (D$_{6h}$ symmetry) for the four leading superconducting pairing gaps with corresponding eigenvalues ($\lambda$). Notably, the non-interacting Hamiltonian employed here incorporates the Ising SOC.  Additional parameters used are $U = 1.0$ and $J = U/4$.}
    \label{fig_nesting_gap_with_soc}
\end{figure}

Unlike before in Figure~\ref{fig_nesting_gap_with_out_soc} we present the solutions of the linearized gap equation for a Fermi surface with D$_{6h}$ symmetry and a pairing interaction calculated from a non-interacting Hamiltonian that takes into account of Ising SOC, thereby reducing its symmetry to D$_{3h}$ symmetry.  Here also the system gets driven towards a dominant spin-triplet pairing instability.

Our main work incorporated the effects of Ising SOC within both the non-interacting Hamiltonian and the Fermi surface. This comprehensive treatment led to results that differ at the subdominant level of the order parameter.  In the main work, the leading subdominant order parameter was identified as a (s+f) singlet-triplet mixture, whereas here, we get an (i+f) singlet-triplet mixture.

\end{widetext}

\end{document}